\documentclass[useAMS,usenatbib,usegraphicx,a4]{mn2e}
\usepackage{times}
\usepackage{url}
\usepackage{epsfig}
\usepackage[flushleft]{threeparttable}
\usepackage{amssymb}
\usepackage{amsmath}
\usepackage{pdfpages}
\usepackage{todonotes}
\usepackage{array}
\usepackage{comment}
\usepackage{caption}
\usepackage{subcaption}
\usepackage[export]{adjustbox}

\def\apjl{ApJL}

\def\apjs{ApJS}

\def\araa{ARAA}
\def\aj{AJ}
\def\aap{A\&A}

\title[SkyMapper colours of Seyfert galaxies]{SkyMapper colours of Seyfert galaxies and Changing-look AGN -- II. Newly discovered Changing-look AGN \\}

\author[Hon et al.]{Wei Jeat Hon$^1$, Christian Wolf$^{2,3}$, Christopher A. Onken$^{2,3}$, Rachel Webster$^1$, 
\newauthor Katie Auchettl $^{1,4,5}$. \\
$^1$School of Physics, University of Melbourne, Parkville, Victoria 3010, Australia \\
$^2$Research School of Astronomy and Astrophysics, Australian National University, Canberra ACT 2611, Australia, E-mail: christian.wolf@anu.edu.au\\
$^3$Centre for Gravitational Astrophysics, Australian National University, Canberra ACT 2611, Australia\\
$^4$ARC Centre of Excellence for All Sky Astrophysics in 3 Dimensions (ASTRO 3D)\\
$^5$Department of Astronomy and Astrophysics, University of California, Santa Cruz, CA 95064,
USA
}

\begin{document}

\date{draft \today}
\maketitle


\begin{abstract}
Changing-look Active Galactic Nuclei (CLAGN) are AGN that change type as their broad emission lines appear or disappear, which is usually accompanied by strong flux changes in their blue featureless continuum. We search for Turn-On CLAGN by selecting type-2 AGN from the spectroscopic 6dF Galaxy Survey, whose photometry, as observed over a decade later by the SkyMapper Southern Survey, is consistent with type-1 AGN. Starting from a random sample of 235 known type-2 AGN we select 18 candidates and confirm 13 AGN to have changed into type-1 spectra; observations of an incomplete sample reveal nine further Turn-On CLAGN. While our search was not intended to reliably discover Turn-Off CLAGN, we discover four such cases as well. This result suggests a Turn-On CLAGN rate of 12\% over $\sim$15 years and imply a total CLAGN rate of $\sim$25\% over this period. Finally, we present observations of AGN that are atypical for the CLAGN phenomenology, including J1109146 - a CLAGN that did not appear as an AGN in 6dFGS; J1406507 - the second reported Changing-look NLS1; and J1340153 - a CLAGN with a change timescale of three months.
\end{abstract}
\begin{keywords}
galaxies: active -- galaxies: Seyfert -- quasars: general -- quasars: emission lines
\end{keywords}

\section{Introduction}\label{intro}
The cause of Changing-look Active Galactic Nuclei (CLAGN) is one of the current mysteries among AGN. AGN are in the cores of galaxies and are powered by an accreting supermassive black hole that generates copious amounts of radiation across the entire electromagnetic spectrum \citep[e.g,][]{BH1, BH2}. This strong source of luminosity, coupled with the physical dynamics between the accretion disk and out-flowing gasses, results in Broad and Narrow Emission Lines (BEL/NEL), which are signature characteristics in an AGN spectrum \citep[e.g.,][]{BH3, BH4}. Briefly, AGN with BEL are known as type-1 AGN, while those without are known as type-2.

Minor stochastic variation in continuum and emission line fluxes are common in AGN \citep[e.g.,][and references therein]{BH5}. On the other hand, changes that would involve a significant variation in continuum flux and the disappearance or appearance of BEL are also possible. The latter would effectively change the AGN type from type-1 to type-2 and vice versa. Such variation follows the viscous timescale that is predicted to be on the order of $100$ to $1,000$ years \citep[][also see examples in \citealp{macleoda}]{BH6, MRI}. 

Surprisingly, CLAGN are observed to exhibit large changes on unexpectedly short timescales, ranging from just a few months to decades. About $30$ cases\footnote{The actual number of CLAGN is not well-defined as the definition of CLAGN in Seyfert galaxies varies between authors.} of CLAGN among Seyfert galaxies have been observed as of December 2020 \citep{TnOCL, PnPCL, CohenCL, BnKCL, ShappeeCL, DenneyCL, RuncoCL, husemann, okynCL, sternCL, WolfCL, trakCL, HonCL}, as well as 54 cases of CLAGN among quasars \citep[also known as CLQ:][]{lamassaCL, ruanCL, macleoda, macleodb, yangCL}. It has been noted that some CLAGN could fit within the timescale of a variable obscuration event, where passing dust clouds cover the line-of-sight to the emission region of the BEL, resulting in a decrease and reddening of flux \citep[e.g.,][]{risaliti09}. However, the observed changes in the H$\beta$ BEL can often not be explained by obscuration alone \citep{macleoda, marinnhutse, ShengCL}. Instead, the more widely accepted cause for CLAGN is a variation in the accretion rate.

Crucially, the expected causes of variations in accretion rate follow the viscous timescale, which is a few orders of magnitude too long for the observed timescale of CLAGN. Either our current models and understanding of accretion disk physics is incomplete, or there are unique dynamical events behind the Changing-look phenomenon. One idea involves the disruption of the accretion disk, suggested by \cite{husemann}, \cite{ruanxray} and \cite{Scepi}. While these are promising models, they are case-specific, rather than generic explanations. The most significant bottleneck to CLAGN studies is the lack of multi-wavelength data, especially during the change. Thus many physical properties of the CLAGN phenomena still remain unconstrained. This includes the changing timescale; the fraction of CLAGN among AGN; the overall profile and characteristic of the light curve around the transition; the behaviour of the spectral energy distribution during the change. The study of CLAGN phenomena will benefit from the next generation of high cadence telescopes and surveys such as the 4-metre Multi-Object Spectroscopic Telescope\footnote{https://www.eso.org/sci/facilities/develop/instruments/4MOST.html} (4MOST) and the Legacy Survey for Space and Time\footnote{https://www.lsst.org/} (LSST). These next-generation telescopes will facilitate a more complete and systematic discovery of CLAGN. In the meantime, we aim to develop effective detection algorithms and prepare for the forthcoming larger samples, which will provide a basis for developing a physical understanding of the phenomenon.

\cite{macleodb} provided statistical evidence to suggest that CLAGN are preferentially found among AGN with lower luminosity and lower black hole mass. Intuitively, smaller black holes will have smaller accretion disks that are expected to be more easily disrupted by clumpy accretion events. Large-scale systematic searches of the Sloan Digital Sky Survey (SDSS) have been designed to target quasars for CLQ \citep{macleoda, macleodb, yangCL}, and the resulting population rate of CLAGN among their total AGN sample is $<0.1\%$. In comparison, Seyfert galaxies are lower luminosity AGN. They are mainly observed at lower redshift, allowing for a larger range of luminosities to be covered by a survey. As expected, the population rates predicted for CLAGN Seyfert galaxies is significantly higher, at 1 to 5\% \citep{RuncoCL, WolfCL, HonCL}. However, these studies are each small-scale searches of $\sim100$ AGN. Therefore, Seyfert galaxies are still a relatively unexplored and attractive area to search for new CLAGN.

As discussed in \cite{HonCL}, the selection method for CLAGN employed by the systematic searches in SDSS is only applicable to quasars and are therefore incompatible with Seyfert galaxies. To briefly summarise, the selection method utilises SDSS \textit{g} and the photometric light curve as an indicator of accretion disk activity, where objects with $>|0.5|$ continuum magnitude variations are considered potential CLAGN. This method is effective for quasars as the photometric magnitude accurately represents the state of the AGN no matter the aperture size, since the host galaxy contribution is not significant and the AGN appears point-like. Aperture photometry fails for Seyfert galaxies that are situated closer due to the extent and contribution of the host. For Seyfert galaxies, there will always be mixing from the stellar component that depends on the aperture size and seeing conditions during the photometric measurement. The best example of an isolated core in a Seyfert galaxy used a 0.1 arcsec aperture \citep[e.g.,]{hst01} and photometry using a ground-based telescope with this small aperture size is not available. While careful modelling and subtraction of the stellar component is possible \citep[e.g.,]{mcelroy}, such techniques cannot be easily scaled to a large scale CLAGN search. 

In \citet[][hereafter, Paper~I]{Wolfsky}, we presented a method that accounts for the mixing of light from the host galaxy, utilising the SkyMapper Southern Survey \cite[SMSS,][]{skymapper2}. This method comprises a single epoch of $u-v$ colour that is used to differentiate between quasars and galaxies. The \textit{u} and \textit{v} bands are centred at 3\,490 and 3\,840\AA, spanning 420 and 280\AA\, respectively. In this wavelength region, AGN at $z < 0.1$ will have a blue $u-v$ colour from the accretion disk's power-law continuum, while galaxies will have a red $u-v$ colour from the drop off in the stellar continuum. This results in a clear separation in $u-v$ colour of these two populations. This method was then applied to a spectroscopic sample of AGN from the Six-degree Field Galaxy Survey \cite[6dFGS,][]{6dfgs}, and highlighted a subset of candidate CLAGN to follow-up.

In a Seyfert galaxy, the mixing of AGN and its host galaxy results in a blend in colour instead of a clear separation (see figure 6 in Paper~I). Seyfert galaxies of type-1 with observable BELs would naturally have a stronger AGN component and hence occupy the colour space closer to pure AGN. Conversely, Seyfert galaxies of type-2 without BELs would occupy the colour space closer towards galaxies. In Paper~I, we derive a cut in colour space that selects for type-1 AGN with high purity. Then, applying this photometric cut to a sample of AGN with past spectra, a Changing-look candidate would be those with contradicting spectral types. e.g., a Turn-On CLAGN candidate will have a type-2 AGN spectrum, but a photometric colour that indicates a type-1 AGN. In Paper~I, this method consistently selects known CLAGN within a restricted sample of 6dF Galaxy Survey (6dFGS) sources. In this paper we extend the 6dFGS sample and follow-up on all the candidates.

For our study, we define CLAGN as an AGN with a strong BEL variation, or in other words, a significant change in AGN type determined spectroscopically. These AGN types were first introduced by \cite{osterbrock}, and include types 1, 1.2, 1.5, 1.8, 1.9 and 2. We adopt the quantitative classification scheme by \cite{winkler}, where 
\begin{align}
    R = \dfrac{flux(\mathrm{total\ H}_{\beta})}{flux([O{\sc III}])} \label{Rval}
\end{align}
defines type-1, 1.2, 1.5, and 1.8 with $R$ values $R>5.0$, $2.0<R<5.0$, $0.333<R<2.0$, $R<0.333$, respectively. type-1.9 has no detectable H$\beta$ broad emission line, while type-2 has no detectable H$\beta$ or H$\alpha$ broad emission lines. From this convention, CLAGN are those that change between the different type categories: 1/1.2/1.5 and 1.8/1.9 and 2. 

We also consider Low-ionization nuclear emission-line region (LINER) galaxies in this paper. LINERs are AGN that are dominated by narrow emission lines of low ionisation such as [S{\sc II}]. Changing-look LINERs have been reported by \cite{clliner}, where LINERs were subsequently found to exhibit BELs. However, the measured flux changes of the broad lines are similar to CLAGN in the main AGN types, so within the scope of this paper we will not place a strong emphasis on the distinction of LINERs and will group them with type-2 AGN. We adopt a flat $\Lambda$CDM cosmology with $\Omega_\Lambda=0.7$ and $H_0=70$~km~s$^{-1}$~Mpc$^{-1}$. Optical magnitudes are in the AB system while mid-IR magnitudes are in the Vega system.

The paper is structured as follows: Section 2 describes the candidate selection and focuses on the data acquisition. Section 3 presents our discovered CLAGN and discusses the implications of the observed spectra for our search method. Section 4 discusses the statistics of our search method and ways to improve it. Section 5 is a summary of the paper. 

\section{Candidate Selection}
Our methodology for CLAGN searches in this paper is as follows. First, we define a sample of AGN within 6dFGS \citep{6dfgs}. Secondly, we apply the selection method described in section \ref{intro} and Paper~I to acquire candidates. Finally, we follow up the candidates with optical spectroscopy. The main difficulty of this method is that AGN in 6dFGS are not classified. Thus, we are currently developing a complete 6dFGS AGN catalogue for this purpose, and here we will use the preliminary results from this catalogue.

\subsection{AGN sample from 6dFGS and SMSS}\label{method:sample}
6dFGS is a near-complete low-redshift (median redshift of 0.053) spectroscopic survey containing 125,071 galaxies \citep{6dfgs}. 95\% have reliable redshifts and 98\% have a SkyMapper Southern Survey (SMSS) counterpart within a 2 arcsec matching radius. We utilise data from the third data release (DR3) of SMSS. The data products in DR3 are the same as DR2 \citep{skymapper2} with an increase in data volume and sky coverage. DR3 provides Main Survey exposures from all six-SkyMapper-bands covering three quarters of the Southern Hemisphere.

From Paper~I, we require SMSS matching sources to have at least one Main Survey \textit{u} and \textit{v} image with a photometric measurement of good quality (\textit{FLAGS}$<4$ and \textit{NIMAFLAGS}$<5$). The SkyMapper node of the All-Sky Virtual Observatory\footnote{\url{https://skymapper.anu.edu.au}} contains a copy of the 6dFGS catalogue that is crossed-matched to the SMSS DR3 \textit{photometry} table using the unique \textit{object\_id} and its counterpart in the 6dFGS table, \textit{dr3\_id}. The \textit{photometry} table contains one row for every single detection of an object in a unique image and can be joined with the \textit{images} table using the \textit{image\_id}. The entry \textit{exp\_time} in this table is required to be 100~sec, corresponding to the Main Survey exposure. SMSS photometry measurements are spread out over several years, and we choose to average the available photometry data per year. We estimate the annual photometry errors by the half-range of the magnitudes within the bin. 

In total, 27,886 sources from 6dFGS were matched to SMSS $u$ and $v$ band main survey images. As the 6dFGS catalogue lacks spectral classifications, and there are no existing complete AGN catalogues of the Southern sky, we use preliminary results from a new classification algorithm (Hon et al., in preparation). Briefly, the algorithm removes the continuum with an iterative smoothing and 3-sigma clipping process, then selects emission line-like features with an adaptive sigma cut, estimating the redshift based on the selected features and matching it to the redshift provided by 6dFGS. Spectra with correctly estimated redshifts are likely to contain genuine emission lines. Currently, this algorithm has a $>$70 percent completeness in AGN selection when compared to the Milliquas catalogue \citep{milliquas}. This selects 800 type-1-like (includes type-1, 1.2 and 1.5) AGN and 269 type-2-like (includes type-1.8, 1.9, 2 and galaxies) AGN\footnote{The selection for type-2 AGN is significantly incomplete compared to type-1-like AGN, due to difficulty separating type-2 AGN from galaxies. We expect type-2 AGN will outnumber type-1 once we have automated line analysis in place.}.

From the total sample size of 1,069, we randomly chose 235 AGN as a test sample for this paper including the sources selected in Paper~I (denoted Paper~II sample, P2S). We will present the completed 6dFGS AGN algorithm and fully describe the remaining sources (including more type-2 AGN) in future papers. We perform line fitting on the type-1-like AGN in P2S to determine the type classification. For the type-2-like AGN, we measure the N{\sc ii} and S{\sc ii} line fluxes for use in BPT classifications to distinguish between type-2 AGN and starforming galaxies \citep{1981PASP...93....5B, BPT_ref}. In particular we used this BPT equation:
\begin{align}
    \mathrm{log}([OIII]/H\beta) &= \dfrac{0.61}{(\mathrm{log}([NII]/H\alpha)-0.05)} + 1.3
\end{align}

\subsection{Line Fitting}\label{linefitting}
Line fitting algorithms are used in order to measure line ratios to classify AGN type. We use the line fitting code PyQSOFit\footnote{\url{https://github.com/legolason/PyQSOFit}} written by \cite{pyqsofit}, but we modified it to fit skewed Gaussians and allow for absorption lines\footnote{A public version of the code can be found here: \url{https://github.com/JackHon55/PyQSOFit\_SBL}}. PyQSOFit performs a continuum fit with a low order polynomial and an optional Fe complex template, which is significant for many Seyfert galaxies. Emission lines are then fitted with skewed Gaussians in small regions around the H$\beta$ and H$\alpha$ emission lines, $4640-5100$\AA\ and $6400-6800$\AA\ respectively. 

The initial input parameters for PyQSOFit include the name of the spectral line to fit, rest wavelength position, allowed velocity offset, standard deviation, and the weighting of the spectral line. These are fitted and translated into centroid wavelength, Full Width at Half Maximum (FWHM), Skewness, equivalent width, and total line flux. PyQSOFit is also able to calculate these fitting properties for multiple-Gaussian line decompositions, except for skewness, which would be highly degenerate with multiple components. Emission lines are then defined by positive initial weighting and will have positive area, while absorption lines will be characterised by negative initial weight and area. Narrow lines are restricted to have a FWHM $<1000$ km~s$^{-1}$, intermediate lines\footnote{Narrow line Seyfert galaxies commonly will have emission lines of this width range, but our sample selection already excludes them. Line fitting that results in intermediate widths therefore marks those that require revision and a more careful fit.} are restricted to $1000 <$  km~s$^{-1}$ FWHM $<2000$ km~s$^{-1}$, and broad lines have $FWHM > 2000$ km~s$^{-1}$. Uncertainties in the fitting parameters are determined in PyQSOFit using a Monte-Carlo chain to perform bootstrapping by randomly tweaking the input spectrum according to the pixel variance.

\subsection{SMSS Colour Cuts}\label{method:srcut}
Paper~I outlines the derivation of the selection cut in detail. Probable type-1-like Seyfert galaxies (i.e., types 1, 1.2 and 1.5) are selected from SMSS at redshifts $<0.1$ with the following equation:
\begin{align}
    u_{5''} - v_{5''} < 0.15 - \rm max (0, 0.3 \times (u_{0,5''} - 17.25)) \label{cut}
\end{align}
where $u_{5''}$ and $v_{5''}$ are the 5 arcsec-aperture magnitudes taken annually. SMSS aperture magnitudes are corrected for aperture losses based on the per-image PSF (for more detail see \cite{skymapper2}). $u_{0,5''}$ is the 5 arcsec-aperture annual magnitude corrected for extinction. Extinction correction is applied using a \cite{F99} extinction law, with dust maps from \cite{SFD98} and a $u$-band absorption coefficient of 4.294 from \cite{WolfCL}. The $u-v$ criteria is redefined as a Selection Rule (SR):
\begin{align}
    SR = (u-v)_{5''} - 0.15 + {\rm max} (0, 0.3 \times (u_{0,5''} - 17.25)) \label{srcut}
\end{align}
which is simply a rearrangement of Eq. \ref{cut}, where objects with an SR $<0$ are selected as potential type-1-like AGN. 

The purity and mixing of objects selected by this method are discussed in Paper~I and variations in the purity with the new spectroscopic data is discussed in Section~\ref{disc:purity}. A type-2-like Seyfert galaxy as observed in 6dFGS spectrum (i.e., types 1.8, 1.9, 2) that has SR$<0$ would currently be a probable type-1-like Seyfert galaxy and therefore a Turn-On CLAGN candidate. We do not claim that all type-1-like Seyfert galaxies have SR$<0$, rather the probability of a type-2-like Seyfert galaxy having SR$<0$ is low. 

In Paper~I, we did not derive a method for selecting Turn-Off CLAGN candidates since there was no photometric cut that would select for type-2 AGN with high purity. Such a selection cut could appear once the spectral typing has been updated with spectroscopic observations. In order to probe for this, we decided to consider any type-1, 1.2 and 1.5 objects with SR$>0$ as ``Turn-Off CLAGN'' candidates, keeping in mind that the probability of these being actual CLAGN would be low.

Finally, we apply the SR selection to the P2S. 18 Turn-On and 24 Turn-Off candidates were chosen for follow-up spectroscopy. Additionally, we recovered NGC 2617, a known Turn-On CLAGN \citep{ShappeeCL}, as well as Mrk 1018, a known CLAGN that has recently turned off \citep{husemann}.

\subsection{Wide-Field Spectrograph (WiFeS) Follow-up Observations and Data Reduction}
The Wide Field Spectrograph\footnote{https://rsaa.anu.edu.au/observatories/instruments/wide-field-spectrograph-wifes} \cite[WiFeS;][]{Dopita07,Dopita10} is an integral-field spectrograph at the Australian National University (ANU) 2.3m telescope located at Siding Spring Observatory. It has a field-of-view of 38 x 25 arcsec, with a 1 x 1 arcsec spaxel when using a Y=2 binning read-out. Our follow-up observations use the RT-560 beam splitter with B3000 and R3000 diffraction gratings for the $3200-5900$\AA\  and $5300-9800$\AA\  wavelength ranges, respectively. For one of the spectra shown, a RT-480 beam splitter was used instead. Our observing period consists of multiple observing runs with dates for each source given in the Appendix.

Calibration for the raw data was taken each night, including bias, flats, arc, wire, sky flats and telluric standard star spectra for flux calibration and telluric removal purposes. Data reduction was performed using PyWiFeS\footnote{http://www.mso.anu.edu.au/pywifes/doku.php} \citep{pywifes}, that will calibrate the raw data, remove bad pixels, remove cosmic rays and produce cube files (two spacial dimensions and one wavelength axis). The spectra are then extracted from these cube files through a simple python code that reads off the data in each spaxel. To compare WiFeS spectra to the 6dFGS counterpart, we use a 29-spaxel tiled pattern that best resembles a 7'' extraction aperture. This was chosen to closely match the 6dFGS 6.7'' fibre aperture. For background subtraction, we extract a part of the sky that is isolated from the source using a similarly sized aperture.

The seeing during our observations averaged 2.2 arcsec, with a range from 1.2 to 2.7 arcsec. With a larger seeing, more of the core AGN light is spread outside of the extraction aperture, while more of the host galaxy stellar component leaks into the aperture. However, since the ANU-2.3m telescope and the Schmidt Telescope that the 6dFGS survey used are both located at Siding Spring, similar seeing conditions and mixing ratios are expected for each set of observations.

\subsection{CLAGN determination and discovery}
We determine our CLAGN by line fitting 6dFGS and our WiFeS spectra to calculate their $R$ values. Objects that have $R$ value variations that cross the type categories 1/1.2/1.5 and 1.8/1.9 and 2, are considered CLAGN. From the total of 42 P2S candidates observed, 15 CLAGN were discovered and are listed in Table \ref{MS-TAB}.

The remaining objects that have a type change or significant $R$ value variation, but do not meet the definition for a CLAGN, are considered to be variable BEL sources and are discussed in Section \ref{belvar}. Plots of all spectra can be found in the Appendix with a few examples shown in Figures \ref{CL-EX1} and \ref{CL-EX2}.

6dFGS spectra are not flux calibrated, and for some there is a mismatch in the scaling between the blue and red parts of the spectra. This means comparing the 6dFGS and WiFeS spectra is not straightforward. For the purpose of identifying CLAGN, these issues are less relevant. The $R$ value is unaffected and the main visual cue for a CLAGN lies in the profile around the H$\beta$ region, i.e. the BEL is either present or not. We choose to present the spectra by normalising to the continuum between $5200 - 6200$\AA\ restframe wavelengths. 

In addition, we also measure the H$\alpha$ BEL luminosity from the WiFeS spectra. This is done by fitting a spectrum extracted from an aperture that captures all of the H$\alpha$ flux. We manually chose an aperture for each individual spectrum for this purpose. All of our spectra usually require around 11 to 13 arcsec diameter aperture before we stop observing an increase in H$\alpha$ BEL flux.

\subsection{Serendipitous CLAGN}\label{filler}
As a part of the effort to build the full AGN catalogue within 6dFGS, we also obtained new WiFeS spectra of additional 6dFGS sources. This resulted in 2 Turn-Off and 9 Turn-On CLAGN from 90 observed AGN, most of which are also within P2S but were not selected as candidates. All these are listed separately in Table \ref{MS-TAB}. We also found J1406507, a Turn-On Changing-look Narrow Line Seyfert 1 (CLNLS1), which we have not included with the CLAGN since the physical mechanism behind the event might differ from a CLAGN (see Sec.~\ref{sec:clnls1}).

\begin{table}
\centering
\caption{Details of samples and observed targets. The +1 for the serendipitous CLAGN indicates a CLNLS1 that we chose to count separately as it might have physically different causes.}
\label{samplesummary}
\resizebox{0.4\textwidth}{!}{%
\begin{tabular}{l|ccc}
 & \multicolumn{1}{l}{Sample Size} & \multicolumn{1}{l}{CLAGN} \\ \hline
SMSS DR3 source & 27,886 &  \\
AGN-Like 6dF spectrum & 1,069  &  \\
Paper~II sample (P2S) & 235 &  \\ \hline
\ \  Turn-On Candidate & 18 & 13 \\
\ \  Turn-Off Candidate & 24 & 2 \\
\hline
Serendipitous CLAGN & 90 & 11(+1) \\ 
\hline
\end{tabular}%
}
\end{table}

\section{Results}
The CLAGN discovered from the P2S and the Serendipitous CLAGN are presented in Figures \ref{CL-EX1} and \ref{CL-EX2}, and in the Appendix. Ultimately, the most useful pieces of information are in the past and present spectra of the CLAGN. In addition to the BEL changes, the short wavelength continua of our AGN spectra have also varied by different amounts. We also identified three CLAGN that are distinct from others and discuss them specifically in section \ref{res:notable}. Among those with multiple WiFeS observations, we observed two that changed back and forth, which gives us a changing timescale. Light curves were used to assist in our analysis for specific cases.

\subsection{CLAGN population and changing timescales}\label{res:time}
Our search method uncovered 15 new CLAGN and recovered 2 known CLAGN from the initial 235 in the P2S. In addition, we also found 11 serendipitous CLAGN from the randomly observed 90 AGN. Taken together,  these indicate a population fraction of $\sim10\%$, similar to the rates reported in \cite{RuncoCL, WolfCL, HonCL}. 

Two of our CLAGN were observed to change type multiple times. The first is J0917272, which was a type-1.9 in 6dFGS in Janurary 2005, type-1.5 in December 2017, type-1.8 in July 2019 and a type-1.9 in July 2020 (Figure \ref{CL-EX2}, third panel). The timescale for this CLAGN to change from type-1.5 to type-1.9 is therefore $\sim2.5$ years. 

The second is J1340153 (Figure \ref{CL-EX1} first panel), which was a type-1.5 in 6dFGS in Janurary 2005, a type-1.9 in February 2020 then a type-1.5 in June 2020. The timescale for the latter change is three months, which is surprisingly short and reminiscent of the CLAGN NGC\,7603 discovered by \cite{TnOCL}.

These, and a few others cases in the literature \citep[e.g.,][]{trakCL}, demonstrate that CLAGN are able to change on a monthly to yearly timeframe. Therefore, a significant percentage of CLAGN are likely to remain undetected because they have reverted type by the time they are re-observed (we discuss a potential case in the appendix). The fraction of CLAGN among AGN could therefore be even higher if we search with finer temporal resolution.

\begin{table*}
\centering
\caption{Properties of our CLAGN sample: 6dFGS redshift; WiFeS epoch and AGN type; 6dFGS epoch and type; $L_{H\alpha}$ from WiFeS derived from total H$\alpha$ BEL flux; SMSS $u_{0,5''}$ and SR magnitude, both derived from the annual magnitude. $\dagger$MJD=58165 is a fixed aperture spectrum is from ePESSTO and $L_{H\alpha}$ measurement can not be done. WiFeS type for this object is derived from a 1 arcsec aperture spectrum in order to be comparable to ePESSTO.}
\label{MS-TAB}
\resizebox{\textwidth}{!}{%
\begin{tabular}{lllccccrcr}
\hline
\multicolumn{9}{c}{\textbf{P2S Turn-On CLAGN}} \\ \hline
SMSS Id & \multicolumn{1}{l|}{Name} & \multicolumn{1}{c}{z} & 6dFGS & 6dFGS & WiFeS & WiFeS & \multicolumn{1}{c}{$L_{H\alpha}$} & $u_{0,5''}$ & \multicolumn{1}{c}{SR} \\
 & \multicolumn{1}{l|}{} & & Epoch (MJD) & Type & Epoch (MJD) & Type & ($10^{42} \mathrm{erg}/s$) & (mag) & \multicolumn{1}{c}{(mag)} \\\hline
11771081    & \multicolumn{1}{l|}{J0041321-223838} & 0.06307 & 52879 & 1.9 & 59397 & 1.2 & 0.902 & 17.390 & $-0.065$ \\
504133901   & \multicolumn{1}{l|}{J0130213-460145} & 0.11152 & 53325 & 1.9 & 59027 & 1.5 & 2.127 & 17.163 & $-0.151$ \\
17326535    & \multicolumn{1}{l|}{J0345125-393429} & 0.04320 & 52663 & 1.8 & 59193 & 1.5 & 0.551 & 15.992 & $-0.119$ \\
22653039    & \multicolumn{1}{l|}{J0454341-200507} & 0.07527 & 53000 & 1.9 & 58904 & 1.5 & 0.741 & 17.434 & $-0.112$ \\
32630353    & \multicolumn{1}{l|}{J0613243-290023} & 0.07051 & 52229 & 2.0 & 58783 & 1.2 & 7.146 & 15.663 & $-0.181$ \\
457468024   & \multicolumn{1}{l|}{J0917272-645628} & 0.08600 & 53388 & 1.9 & 58111 & 1.5 & 10.166 & 17.094 & $-0.037$ \\
 & \multicolumn{1}{l|}{} &  &  &  & 58667 & 1.8 & 3.805 &  & \\
 & \multicolumn{1}{l|}{} &  &  &  & 59026 & 1.9 & 2.804 &  & \\
89907492    & \multicolumn{1}{l|}{J1008486-095451} & 0.05725 & 53389 & 1.8 & 59018 & 1.2 & 2.565 & 15.351 & $-0.169$ \\
93052952$^{\dagger}$    & \multicolumn{1}{l|}{J1109146-125554} & 0.02551 & 52672 & 2.0 & 58165 & 1.0 & - & 16.774 & $-0.022$ \\
 & \multicolumn{1}{l|}{} &  &  &  & 59018 & 1.5 & 0.110 &  & \\
203556779   & \multicolumn{1}{l|}{J1824525-432858} & 0.07186 & 52797 & 1.9 & 59019 & 1.0 & 1.967 & 17.390 & $-0.132$ \\ 
299274242   & \multicolumn{1}{l|}{J2007513-110835} & 0.03109 & 52902 & 2.0 & 58668 & 1.5 & 0.180 & 16.675 & $-0.098$ \\
 & \multicolumn{1}{l|}{} &  &  &  & 59019 & 1.5 & 0.032 &  & \\
487504676   & \multicolumn{1}{l|}{J2037599-502334} & 0.06311 & 53286 & 1.9 & 59026 & 1.0 & 0.985 & 17.158 & $-0.179$ \\
2679475     & \multicolumn{1}{l|}{J2309192-322957} & 0.05417 & 52462 & 1.9 & 59026 & 1.5 & 0.391 & 17.449 & $-0.046$ \\
8493273     & \multicolumn{1}{l|}{J2330323-022745} & 0.03322 & 53236 & 1.9 & 58727 & 1.2 & 0.806 & 16.126 & $-0.244$ \\ 
 & \multicolumn{1}{l|}{} &  &  &  & 59026 & 1.2 & 0.764 &  & \\ \hline
 
\multicolumn{9}{c}{\textbf{P2S Turn-Off CLAGN}} \\ \hline
107903088   & \multicolumn{1}{l|}{J1340153-045332} & 0.08654 & 53473 & 1.5 & 58906 & 1.9 & 0.715 & 18.921 & 0.620 \\
 & \multicolumn{1}{l|}{} &  &  &  & 59016 & 1.5 & 1.049 &  & \\
 & \multicolumn{1}{l|}{} &  &  &  & 59308 & 1.5 & 1.546 &  & \\
304230428   & \multicolumn{1}{l|}{J2046446-012208} & 0.02519 & 53207 & 1.5 & 58726 & 1.9 & - & 17.451 & 0.217 \\ 
 & \multicolumn{1}{l|}{} &  &  &  & 59080 & 1.9 &  &  & \\ \hline
 
\multicolumn{9}{c}{\textbf{Serendipitous CLAGN}} \\ \hline
13120033    & \multicolumn{1}{l|}{J0010100-044238} & 0.02937 & 52971 & 1.9 & 58669 & 1.5 & 0.132 & 17.491 & 0.051 \\
13121698    & \multicolumn{1}{l|}{J0014181-051904} & 0.08398 & 52971 & 1.9 & 59016 & 2.0 & - & 18.318 & 0.295 \\
491746524   & \multicolumn{1}{l|}{J0248345-720831} & 0.07572 & 53648 & 2.0 & 58906 & 1.9 & - & - & \multicolumn{1}{c}{-} \\
22635196    & \multicolumn{1}{l|}{J0458403-215931} & 0.03944 & 52930 & 1.9 & 58814 & 1.5 & 0.419 & 18.046 & 0.270 \\
19807012    & \multicolumn{1}{l|}{J0519358-323928} & 0.01258 & 52669 & 1.9 & 58746 & 1.5 & 0.387 & 15.688 & 0.167 \\
31686796    & \multicolumn{1}{l|}{J0612386-354307} & 0.04415 & 52226 & 1.9 & 58903 & 1.5 & 0.255 & 17.526 & 0.112 \\
170151522   & \multicolumn{1}{l|}{J1538448-032248} & 0.02374 & 53089 & 1.8 & 58666 & 1.5 & 0.265 & 15.589 & 0.002 \\
370411405   & \multicolumn{1}{l|}{J1721342-544315} & 0.06234 & 53207 & 1.9 & 59019 & 1.2 & 0.152 & 17.277 & 0.064 \\
216993546   & \multicolumn{1}{l|}{J1839358-354124} & 0.04567 & 52382 & 1.5 & 58668 & 1.8 & - & 17.418 & $-0.043$ \\
 & \multicolumn{1}{l|}{} &  &  &  & 58727 & 1.8 &  &  & \\
295796207   & \multicolumn{1}{l|}{J1949137-184716} & 0.08113 & 52897 & 2.0 & 58222 & 1.5 & 2.326 & 18.197 & 0.004 \\
 & \multicolumn{1}{l|}{} &  &  &  & 58377 & 1.5 & 2.204 &  & \\
 & \multicolumn{1}{l|}{} &  &  &  & 58667 & 1.5 & 1.878 &  & \\
 & \multicolumn{1}{l|}{} &  &  &  & 59016 & 1.5 & 2.028 &  & \\ 
486962794   & \multicolumn{1}{l|}{J1958565-422230} & 0.03203 & 53204 & 1.9 & 59074 & 1.2 & 0.449 & 17.229 & 0.077 \\ 
 & \multicolumn{1}{l|}{} &  &  &  &  &  &  &  & \\
102488645   & \multicolumn{1}{l|}{J1406507-244250} & 0.04570 & 52368 & 2.0 & 59019 & NLS1 &  &  &  \\\hline
\end{tabular}%
}
\end{table*}

\subsection{Correlation between Continuum and Line Changes}\label{res:continuum}
\begin{figure*}
    \centering
    \includegraphics[width=\textwidth]{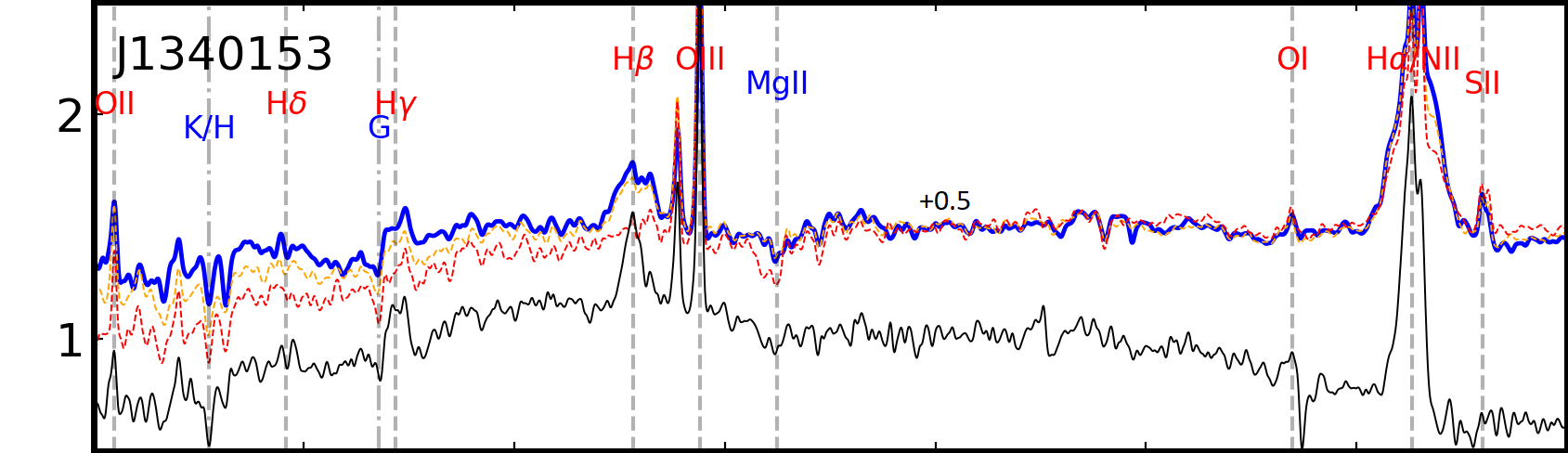}\vspace{-0.2em}
    \includegraphics[width=\textwidth]{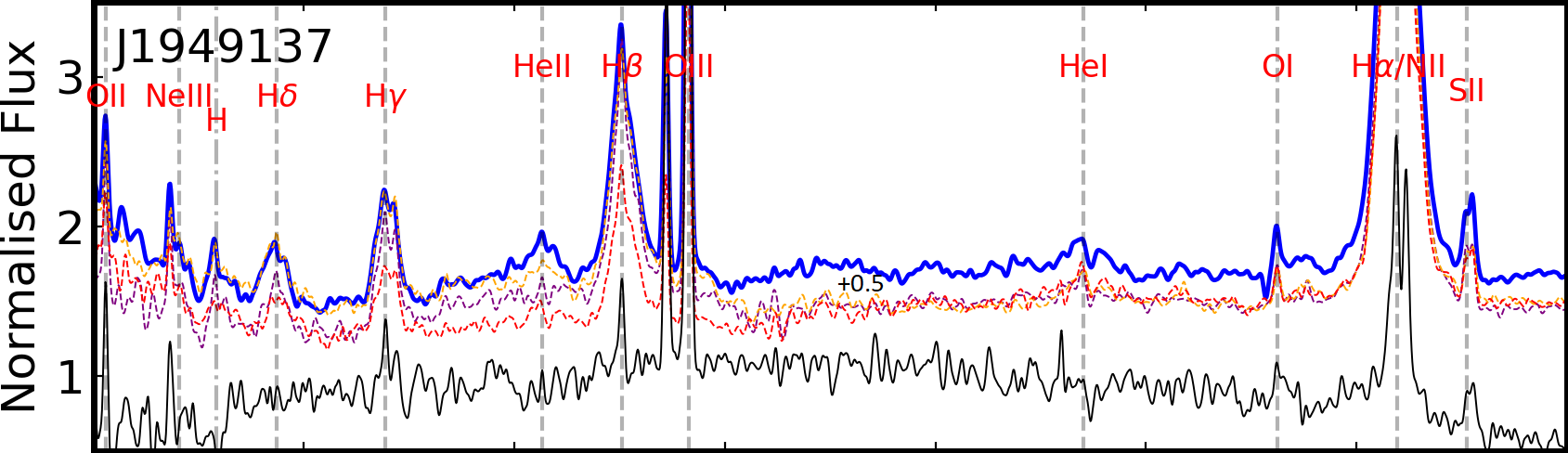}\vspace{-0.2em}
    \includegraphics[width=\textwidth]{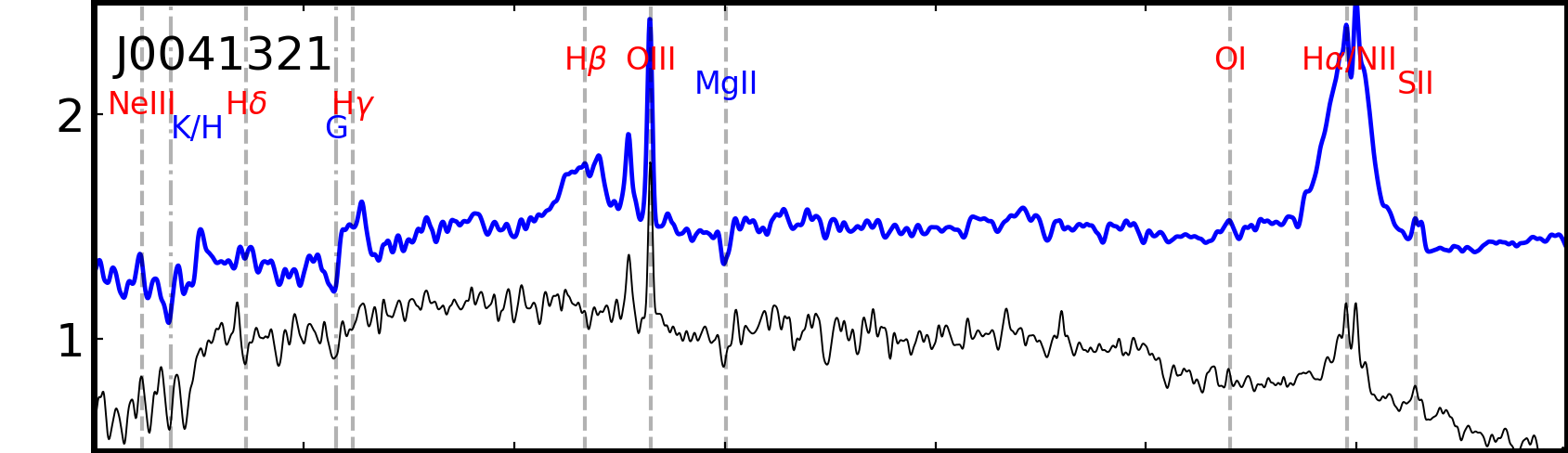}\vspace{-0.2em}
    \includegraphics[width=\textwidth]{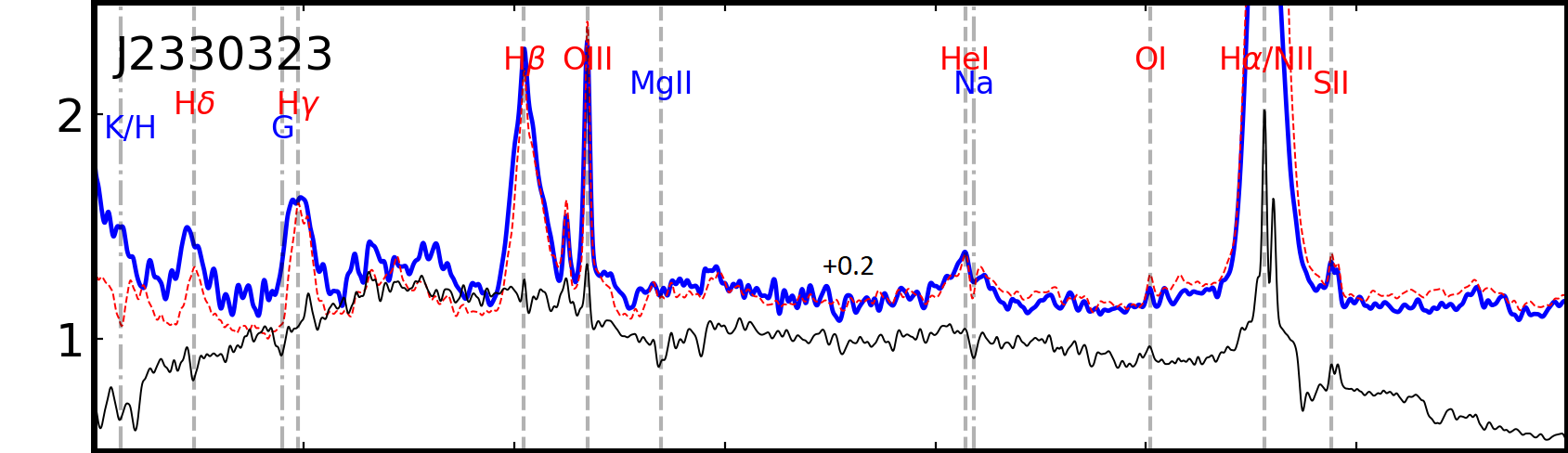}
    \includegraphics[width=\textwidth]{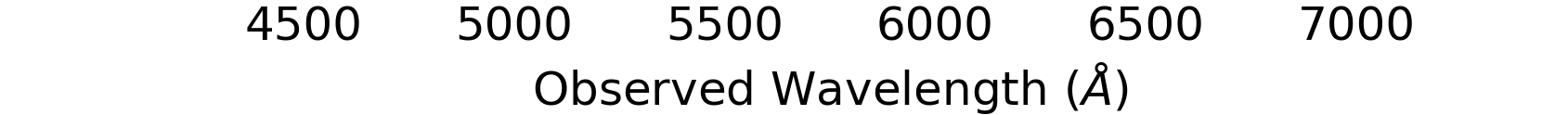}
    \caption{Here we present a few CLAGN that are discussed multiple times in the text. J1340153 is the CLAGN with a 3-month changing timescale. J1949137 has persistent BEL for over two years. J0041321 is an example where the short wavelength continuum remains red in WiFeS, and J2330323 is where the short wavelength continuum turns blue instead. Spectra are flux normalised and presented with the observed wavelength. The blue thick spectra are the most recent WiFeS observation. The black thin spectra are the oldest and from 6dFGS. The dashed spectra are additional observations in between. Vertical dotted lines indicate emission line wavelength location. Some spectra are vertically offset for clarify, where the amount of offset is indicated next to the spectra. The 5578\AA\ skyline has been removed from all spectra. The remaining spectra of our CLAGN are plotted in the Appendix in Figs \ref{CL-APDX1}-\ref{CL-APDX5}}
    \label{CL-EX1}
\end{figure*}

While the AGN is in the Off-state, the short wavelength continuum will naturally be red from the host galaxy. As the CLAGN brightens and BEL appears, the AGN contribution becomes relevant and thus the short wavelength continuum would vary as well. The degree of this change indicates both the luminosity of the AGN and how much the mixing ratio favours the AGN.

e.g., we compare J0041321 ($L_{H\alpha}=0.902\times10^{42} $ erg$/$s, SR=$-0.065$ magnitude) to J2330323 ($L_{H\alpha}=0.806\times10^{42} $ erg$/$s, SR=$-0.244$ magnitude). The luminosity is similar, but the SR value for the latter is much bluer. The profile of the short wavelength continua in WiFeS of these two CLAGN are vastly different, with spectra of J2330323 indicating a mixing ratio that is highly AGN-favoured (see figure \ref{CL-EX1}, third and fourth panel). This example also suggests that the SR magnitude can be used to quantify mixing ratios in Seyfert galaxies as a function of $L_{H\alpha}$. We discuss this aspect in section \ref{disc:purity}.

Within the P2S CLAGN, J0613243, J1008486, J1824525 and J233032 (figures in appendix) have short wavelength continua that transitions from red to blue between the 6dFGS and WiFeS epochs. In comparison, J0041321, J0345125, J0458403, J2007513 and J2037599 either have lesser changes in their short wavelength continua or remain red in their WiFeS spectra. The $L_{H\alpha}$ and SR magnitude of the two groups are median values of $2.266\times10^{42} $ erg$/$s and $-0.175$ magnitude, and $0.646\times10^{42} $ erg$/$s and $-0.105$ magnitude, respectively. This is a large difference in luminosity, but the SR magnitude for the fainter group is still significantly below zero (lowest SR value is $-0.046$ magnitude).

In comparison, almost all of the CLAGN presented in \cite{macleoda, macleodb} have strong red to blue transitions in their short wavelength continua due to the $|\Delta g|>1$ magnitude requirement. This requirement would select only AGN that have mixing ratios that are strongly AGN-favoured, and thus be biased towards AGN that attain high luminosity or those with smaller host galaxy. \cite{yangCL} presented CLAGN that are found by comparing repeat spectroscopy with no magnitude restrictions. Such a search avoids the luminosity bias and indeed their CLAGN have a large variety of short wavelength continuum changes. 

\begin{figure*}
    \centering
    \includegraphics[width=\textwidth]{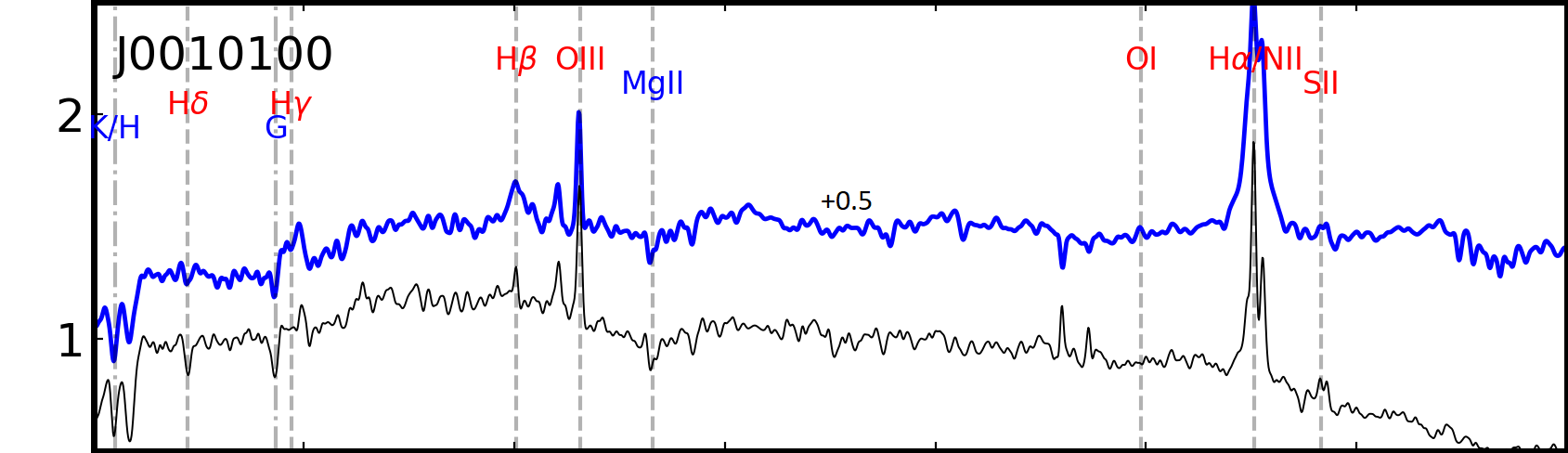}\vspace{-0.2em}
    \includegraphics[width=\textwidth]{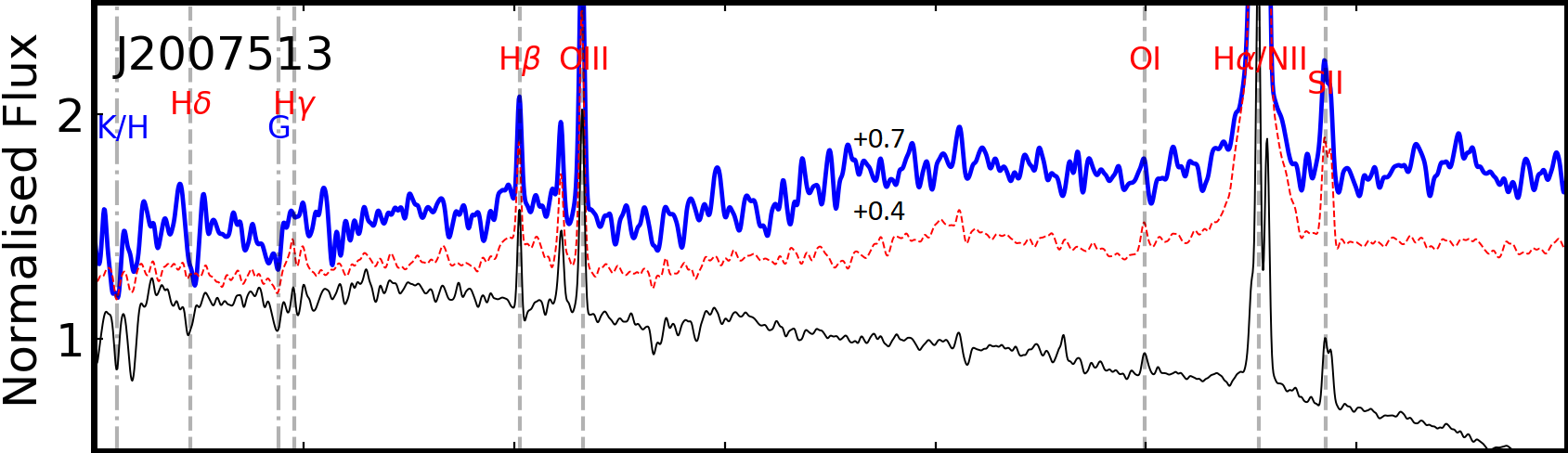}\vspace{-0.2em}
    \includegraphics[width=\textwidth]{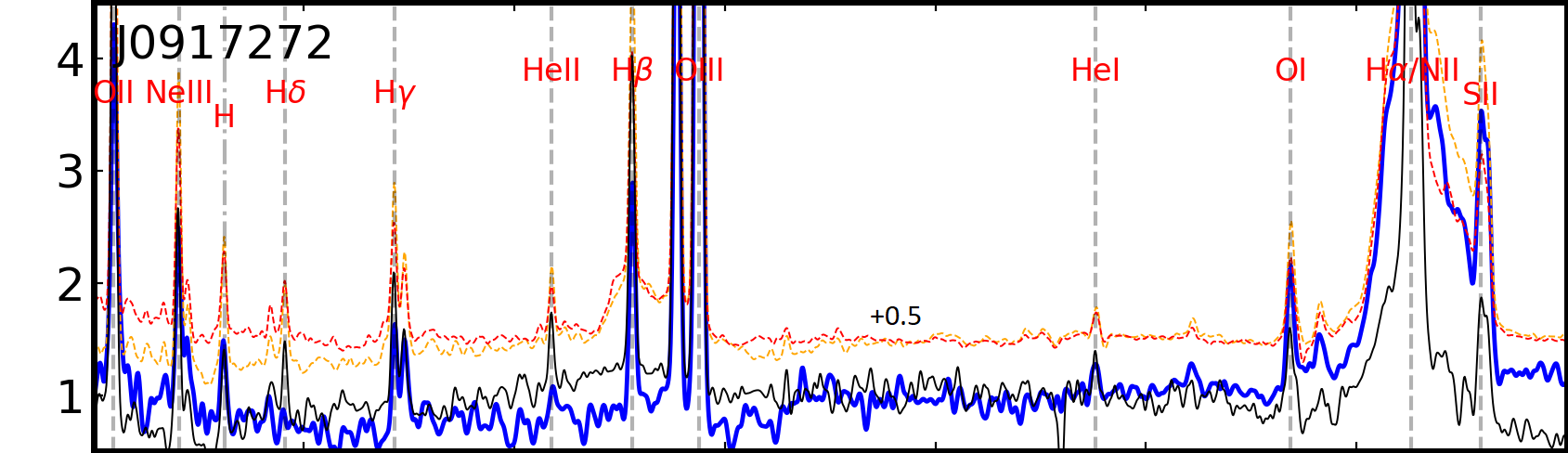}\vspace{-0.2em}
    \includegraphics[width=\textwidth]{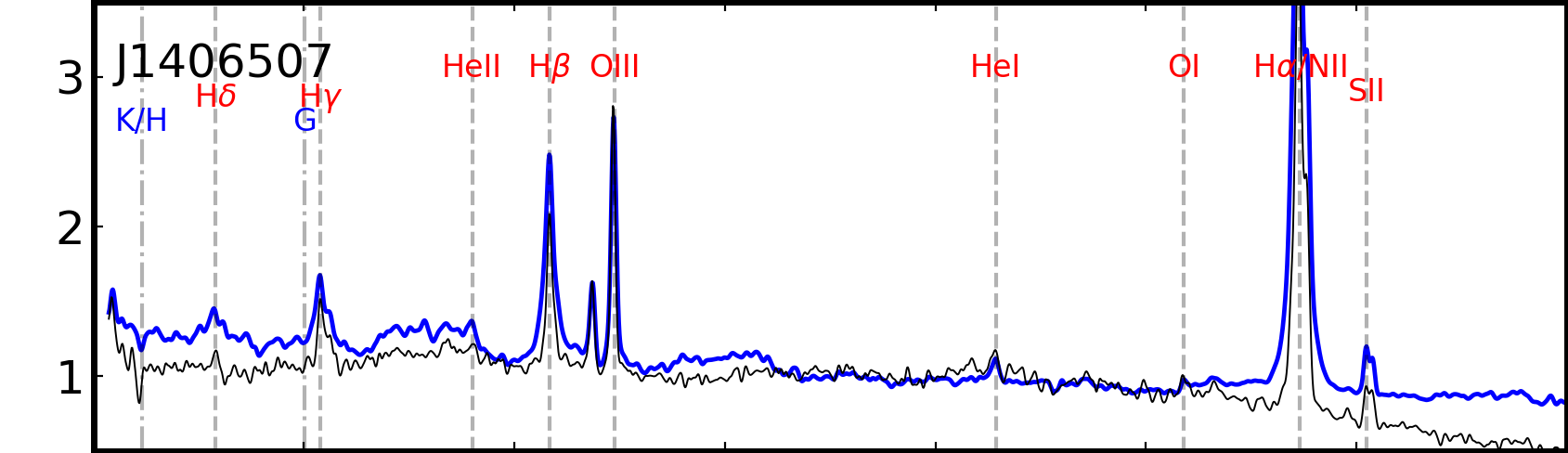}
    \includegraphics[width=\textwidth]{CLAGN-Spec/xaxis-spec.png}
    \caption{Similar to Figure \ref{CL-EX1}. J0010100 is an example of serendipitous CLAGN that we believe is too dominated by the host galaxy for our search method to select. J2007513 is a CLAGN that we caught fading as indicated by WISE W1 magnitude and SR magnitude. J0917272 is a CLAGN that has turned on and off, but has strong blue continuum and NELs. J1406507 is a CLNLS1}
    \label{CL-EX2}
\end{figure*}

However, our method should fail for AGN with a very low contrast against the host galaxy. This is likely the case for the serendipitous CLAGN J0010100, J0458403, J1538448, J1721342 and J1958565 with SR magnitudes that are above 0. The WiFeS spectra (e.g., see J0010100 in figure \ref{CL-EX2}, first panel) have short wavelength continua that remain red. The luminosity dependence of the SR magnitude (discussed in Sec.~\ref{disc:on-success}) also indicates a degree of bias towards higher-luminosity AGN.

\begin{figure}
    \includegraphics[width=0.47\textwidth,right]{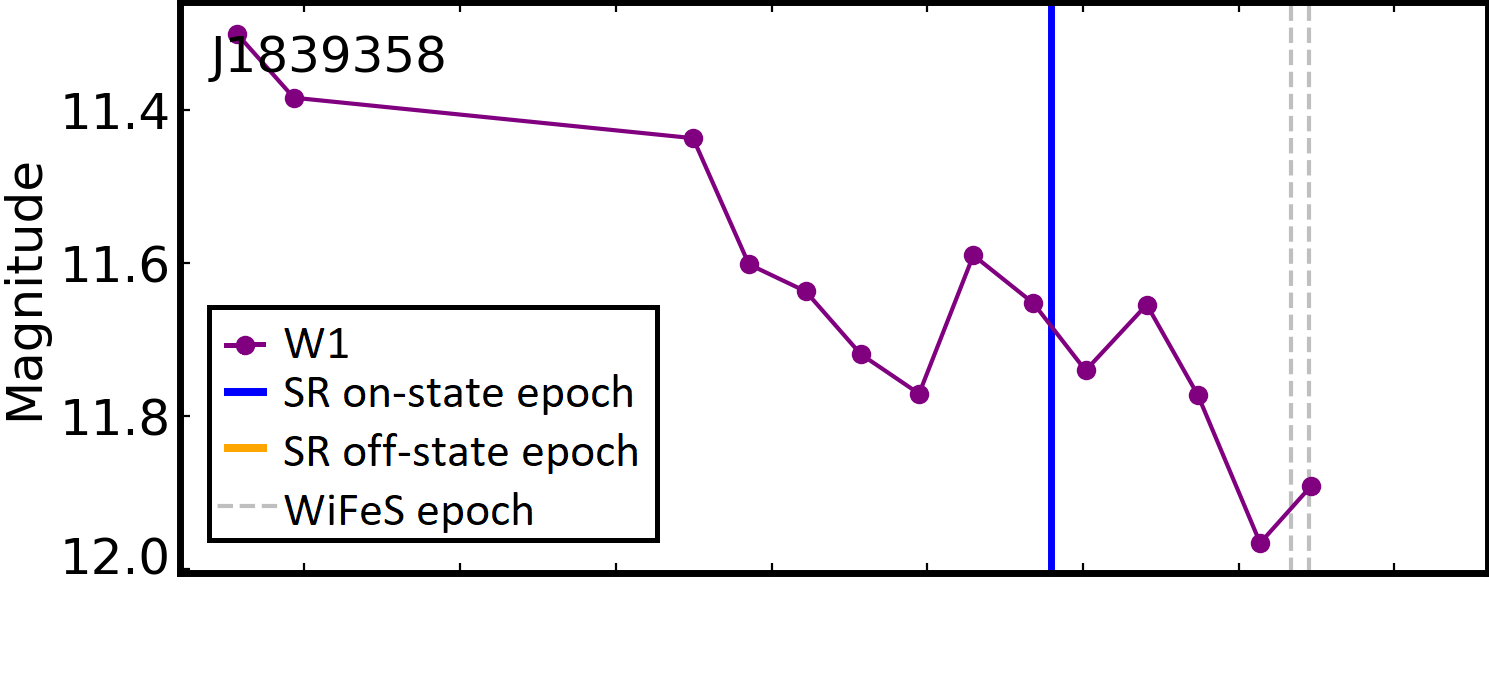}\vspace{-2.45em}
    \includegraphics[width=0.471\textwidth,right]{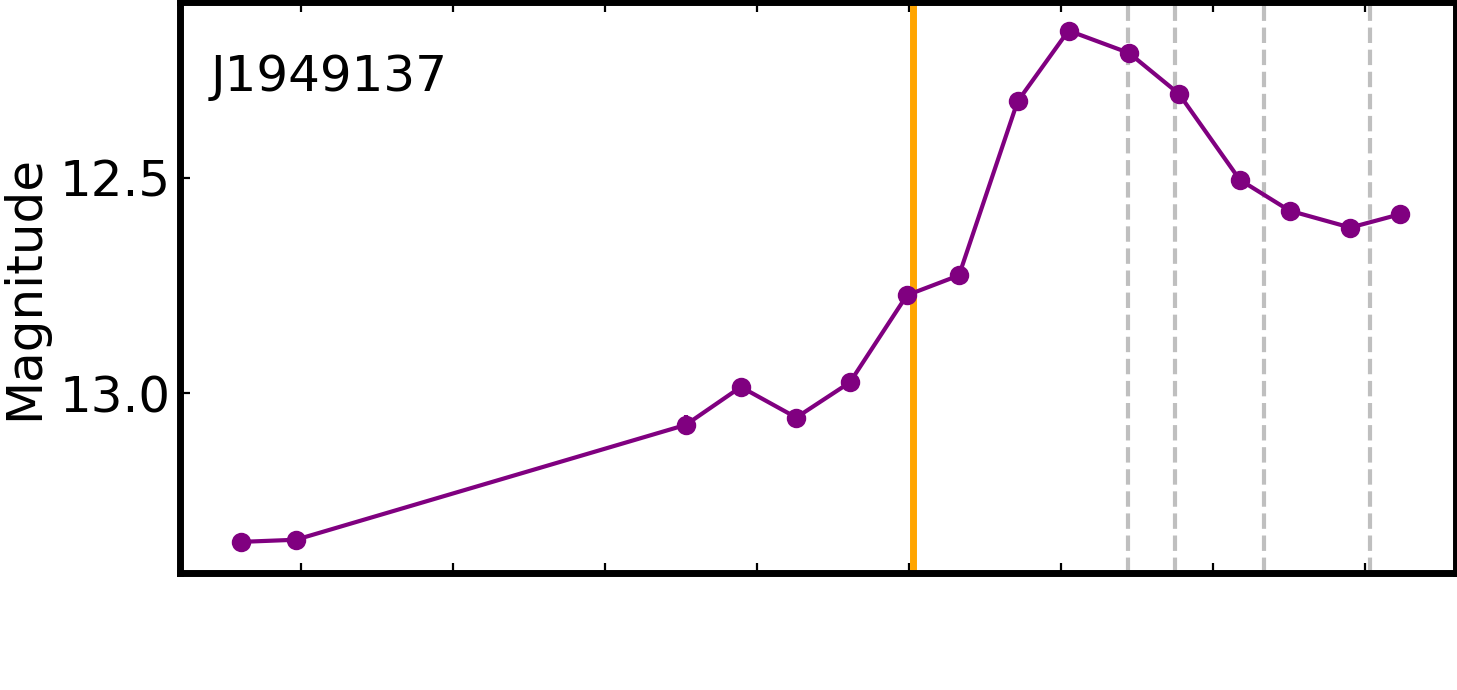}\vspace{-2.46em}
    \includegraphics[width=0.47\textwidth,right]{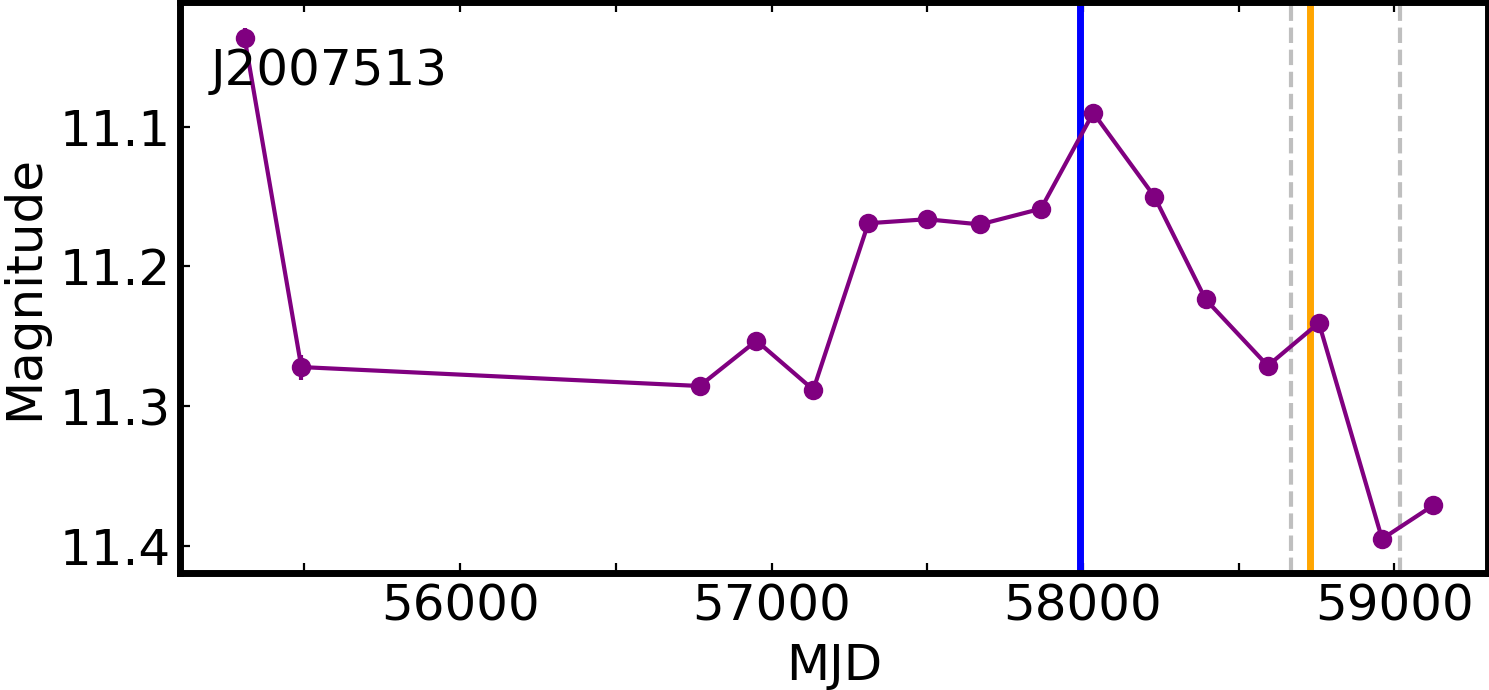}
    \caption{WISE light curves are created with 50 day bin sizes and excluding data that are 2.5 standard deviations away from the median. The times of the spectral observations and the SMSS photometry are also marked.}
    \label{haCL-LC}
\end{figure}

\subsection{Correlation between BEL variation and Mid-IR photometry}\label{res:midir}

The Mid-IR light curve, namely W1 and W2 from the Wide-field Infrared Survey Explorer \citep[WISE,][]{neowise} is known to correlate with CLAGN evolution. This is seen in CLAGN discovered via searching for AGN with magnitude variations in W1 \citep{sternCL, yangCL}. \cite{ShengCL} also report a W1 variation of $>0.4$ magnitude at a $>10\sigma$ level when a Changing-look event occurs. We should therefore expect W1 variations to also correlate with the SR magnitude of CLAGN.

We observe this behaviour in J2007513. This is a CLAGN from the P2S with significant variation in Mid-IR before the WiFeS observation (see figure \ref{haCL-LC}). This is the only candidate with two epochs of SMSS annual photometry, one at MJD=57990 with SR=$-0.098$ indicating a type-1-like AGN, then at MJD=58730 with SR=0.046 indicating a type-2-like AGN. The two WiFeS spectra are taken close to the second photometry epoch, and reflect the change in SR value with a decrease in BEL flux. The first WiFeS spectrum has $R=1.26$ and $L_{H\alpha}=0.180\times10^{42} $ erg$/$s. The second has $R\sim0.6$ and $L_{H\alpha}=0.032\times10^{42} $ erg$/$s.

Given that W1 variation correlates to SR magnitude, we could use this to explain why our selection method could not account for some of our serendipitous CLAGN. We find that this applies to J1839358 and J1949137. J1839358 had W1 dimmed by $\sim0.4$ magnitude between SMSS and WiFeS, while J1949137 had W1 brighten by $\sim0.6$ (see their light curves in figure \ref{haCL-LC}). In these two cases, our selection method did not consider them as candidates solely because their photometry measurements were taken too early. The CLAGN evolved in the period of time between SMSS and WiFeS observation. The existence of J1340153 with a changing timescale of only 3 months makes this explanation highly possible.

We do not observe and can't expect similar behaviour for all CLAGN since the event has to occur within the operational period of WISE. However, we can certainly use WISE light curve in the future to supplement the search method. This will bridge the one- to five-year time gap between SMSS and WiFeS observations in which we should expect significant evolution.

As an aside, there are two cases among our observed sample where we have potentially missed the On-state. This is inferred by a strong dimming in their W1 light curves prior to WiFeS observation. We discuss these two AGN in the appendix.

\subsection{Comments on atypical CLAGN}\label{res:notable}
\subsubsection{J0917272 - CLAGN with extremely strong NEL}
The spectra of J0917272 (see figure \ref{CL-EX2}, third panel) have much stronger NELs relative to BELs when compared to other CLAGN. The H$\beta$ at the strongest and earliest WiFeS spectrum amounts to only a small $R=0.37$, even though $L_{H\alpha} = 10.166\times10^{42}$ ergs~s$^{-1}$ and is the brightest luminosity measured among our CLAGN. The Balmer BEL decreases in flux and the H$\beta$ disappears by the third WiFeS spectrum. The short wavelength continuum also remained blue throughout thr 6dFGS and WiFeS spectra regardless of the presence of H$\beta$ BEL. This is unlike the other CLAGN where the short wavelength continuum either remains red or changes according to the appearance of H$\beta$ BEL. The timescale on which the appearing H$\beta$ BEL vanished is also short (see more in section \ref{res:time}). The behaviour of the continuum and the strong NELs suggests that the host galaxy could have high levels of star formation or perhaps the changes are caused by another process. Close monitoring would be beneficial to understanding its non-standard CLAGN behaviour.

\subsubsection{J1109146 - CLAGN that was not an AGN in the past}
The 6dFGS spectrum of J1109146 does not show strong signs of any AGN activity, with weak NELs that indicate that an AGN was not present long enough to establish emission lines. The All-Sky Automated Survey for Supernovae \citep[ASAS-SN,][]{ASAS-SN}\footnote{\url{https://www.wis-tns.org//object/2018ts}} and the Zwicky Transient Factory \citep[ZTF,][]{ZTF}\footnote{\url{https://lasair.roe.ac.uk/object/ZTF19aabyrnq/}} had flagged this object as a transient with a flare 2.4 arcsec away from the centre, and ePESSTO had taken a spectrum before our WiFeS observation. Since the comparison between WiFeS and ePESSTO will be more useful, we extract the WiFeS spectrum with a 1 arcsec aperture instead, comparable to the 1 arcsec aperture slit used in ePESSTO.

Throughout the three spectra of this object, we see development of the S{\sc ii} lines, and potentially the O{\sc iii} lines as well. The Balmer BELs, most noticeably the H$\beta$ BEL, faded rapidly within two years, with $R=13.42$ in the ePESSTO spectrum but $R=0.936$ in the WiFeS spectrum. These rapid changes and the development of narrow lines, and given that the 6dFGS spectrum indicates that the object might not have been an AGN, leads us to suspect that the physical cause of this CLAGN is different from the rest. We include a discussion on the light curves in the appendix. 

\begin{figure}
    \centering
    \includegraphics[width=0.47\textwidth]{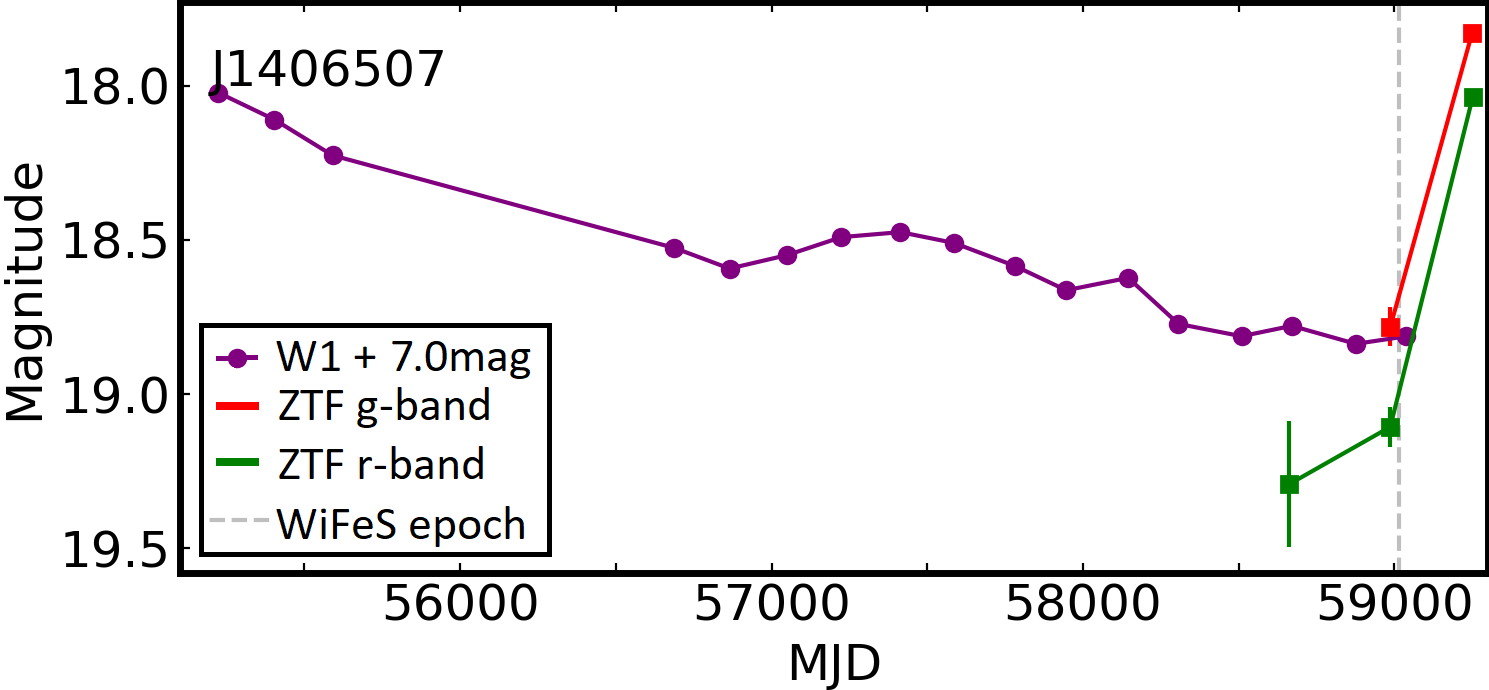}
    \caption{WISE light curve are created with 50 day bin sizes and excluding data that are 2.5 standard deviations away from the mean. W1 magnitude is offset to fit the same scale. ZTF light curves are difference magnitude with 25 day bin sizes.}
    \label{res:clnls1:lc}
\end{figure}

\subsubsection{J1406507 - A CLNLS1}\label{sec:clnls1}
J1406507 is potentially the second reported CLNLS1 with an actual variation to the emission line width. J1406507 is a type-2 in 6dFGS with a H$\beta$ line width of FWHM$\sim500$ km$/$s. The WiFeS spectrum shows a NLS1 with a H$\beta$ FWHM$\sim2400$ km$/$s. While this does not fit the classical FWHM$\leq$2000 km$/$s criteria from \cite{goodrich89}, it is commonly agreed that the line width has no sharp cut off between BELs of AGN and those of NLS1. e.g., \cite{NLS12} increased the threshold to 2200 km$/$s, and \cite{cracco16} allowed for 3000 km$/$s to be inclusive. In addition, there is also the appearance of the Fe emission line complex redward and blueward of the H$\beta$ \citep{NLS11} region, which is a prominent feature of NLS1 and a change that we do not observe in other CLAGN.

Compared to CLAGN, there are not many cases of CLNLS1. A few of the current reported potential cases do not include the appearance of the broad lines, but rather a strong increase in flux. These include PS16dtm \citep{blanchard17}, CSS100217 \citep{drake11} and J1236511 \citep{HonCL}. The only other case with an appearance of broad lines is SDSS J123359.12+084211.5 \citep{macleodb}. Variations observed in PS16dtm and CSS100217 are not believed to be caused by AGN activity and there was limited information about the other two.

When we observed the object with WiFeS, the WISE W1 magnitude had dimmed by almost 1 magnitude since MJD=$55000$ (see figure \ref{res:clnls1:lc}). The ZTF light curve instead shows that the source has brighten by over 1 magnitude just after our WiFeS observation.

\subsection{Comments on AGN with BEL variation}\label{belvar}

\begin{figure*}
    \centering
    \includegraphics[width=\textwidth]{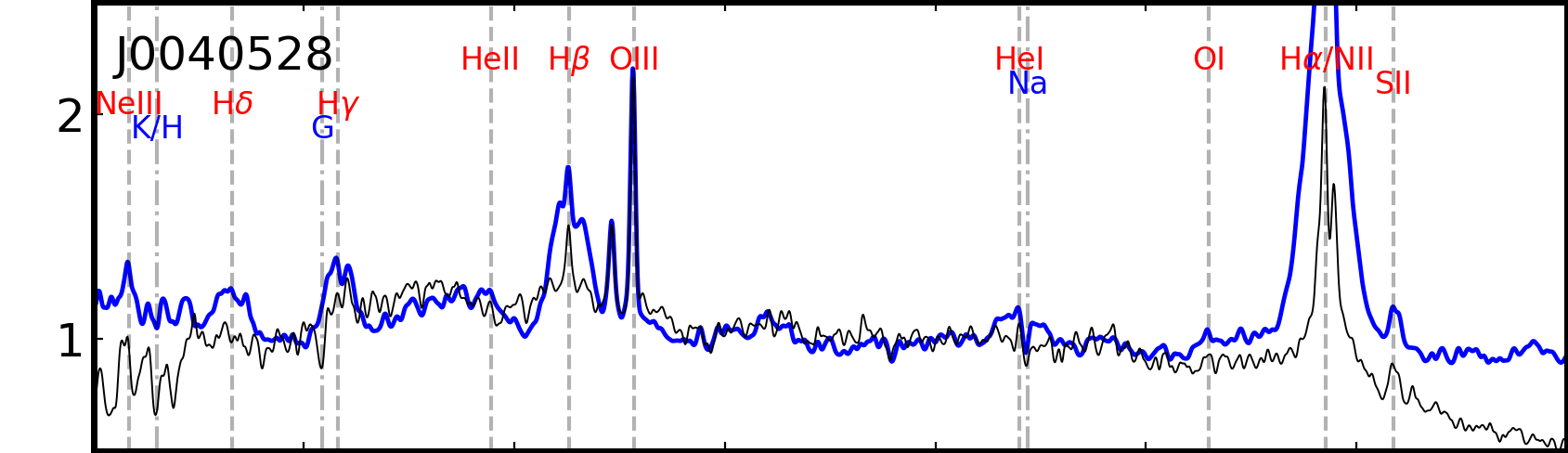}\vspace{-0.2em}
    \includegraphics[width=\textwidth]{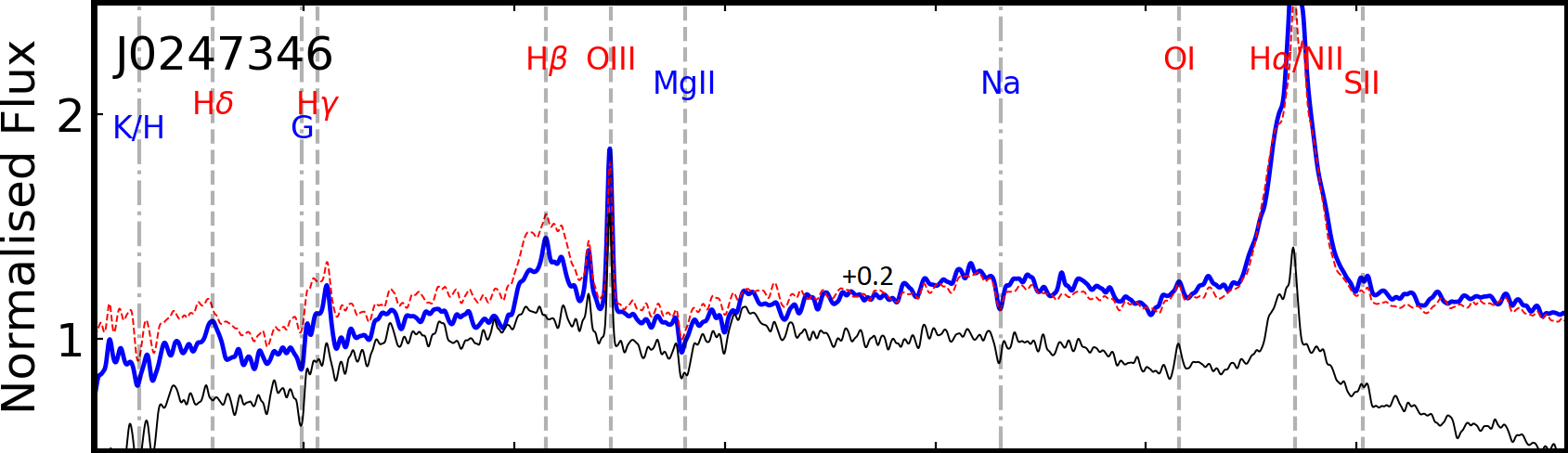}\vspace{-0.2em}
    \includegraphics[width=\textwidth]{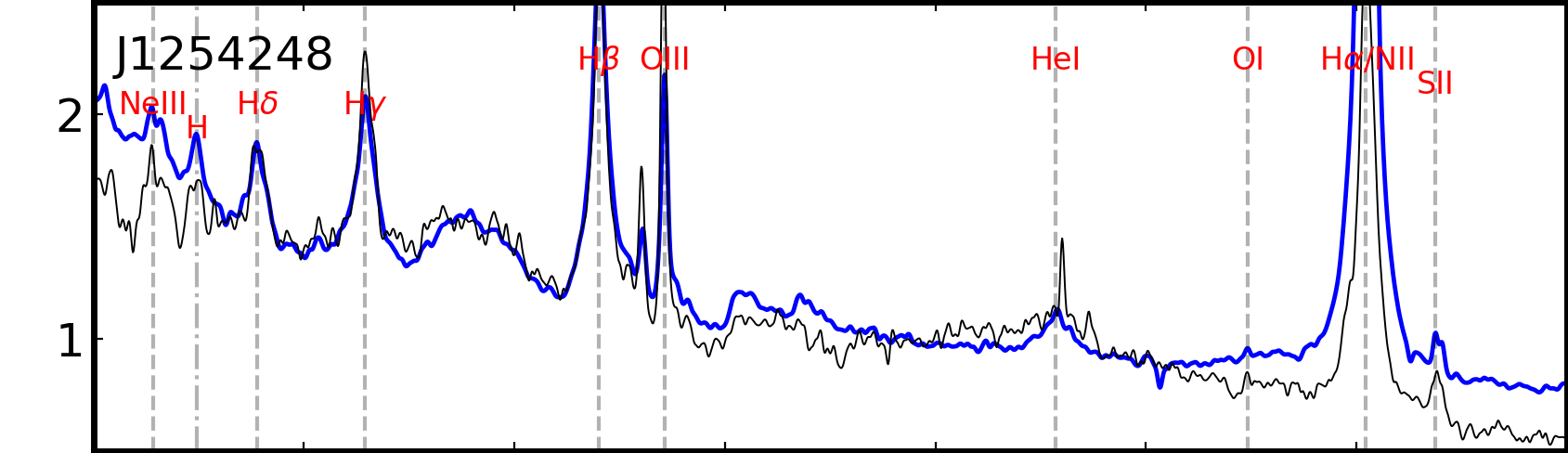}
    \includegraphics[width=\textwidth]{CLAGN-Spec/xaxis-spec.png}
    \caption{Similar to Figure \ref{CL-EX1}, but for the objects with a significant BEL variation. Presented here are examples mentioned in the text.}
    \label{BV-EX}
\end{figure*}

\begin{table*}
\centering
\caption{Similar to Table \ref{MS-TAB}, but providing  $R$ values instead of AGN type to better represent the BEL variations. $R$ values with "?" are measurements where the H$\beta$ BEL is arbitrary due to the continuum profile (see text).}
\label{BV-TAB}
\resizebox{0.75\textwidth}{!}{%
\begin{tabular}{llccccc}
\hline
\multicolumn{7}{c}{\textbf{AGN with Variable BEL}} \\ \hline
SMSS Id & \multicolumn{1}{l|}{Name} & z & 6dFGS Epoch & $R$ & WiFeS Epoch & $R$ \\ \hline
13037273    & \multicolumn{1}{l|}{J0040528-074209} & 0.0551 & 52545 & 1.51 & 59079 & 3.29 \\
14658186    & \multicolumn{1}{l|}{J0247346-201100} & 0.0441 & 53258 & 1.79? & 58783 & 5.31 \\
 & \multicolumn{1}{l|}{} &  &  &  & 59079 & 2.95 \\
17118637    & \multicolumn{1}{l|}{J0313248-350626} & 0.1149 & 52175 & 2.69 & 58905 & 0.50 \\
98109150    & \multicolumn{1}{l|}{J1254248-282632} & 0.0697 & 52822 & 2.63 & 59019 & 4.65 \\
408665904   & \multicolumn{1}{l|}{J1514420-812338} & 0.0690 & 53476 & 0.67? & 58726 & 1.73 \\
 & \multicolumn{1}{l|}{} &  &  &  & 59015 & 2.28 \\
 & \multicolumn{1}{l|}{} &  &  &  & 59346 & 1.57 \\
476888663   & \multicolumn{1}{l|}{J1906088-485027} & 0.0486 & 52762 & 1.42? & 59078 & 4.39 \\
299047923   & \multicolumn{1}{l|}{J1949093-103425} & 0.0240 & 53171 & 4.45  & 59074 & 2.04 \\
4857898     & \multicolumn{1}{l|}{J2117416-020834} & 0.0902 & 52557 & 6.77 & 58668 & 3.13  \\
3799344     & \multicolumn{1}{l|}{J2155318-205133} & 0.0669 & 52876 & 0.6-1.39? & 59078 & 1.26  \\ \hline
\end{tabular}%
}
\end{table*}

There are 9 objects (listed in Table \ref{BV-TAB}) with significant $R$ value variations. Five of them, J0040528 (see figure \ref{BV-EX} first panel), J0313248, J1254248, J1949093 and J2117416, have BEL variations but remain in the type-1, 1.2 and 1.5 category. Therefore, these are not considered CLAGN.

For J0247346 (see figure \ref{BV-EX} second panel), J1514420, J1906088 and J2155318, the presence of a H$\beta$ BEL in the 6dFGS spectra is not visually obvious. Our line fitting tool was unable to provide a definitive result due to the continuum profile around H$\beta$, especially for the extremely broad profiles of J1514420. Therefore, we assume that four objects were type-1.5 AGN in 6dFGS with a BEL flux increase in WiFeS, and do not meet the requirement for a CLAGN. If this assumption is incorrect, then these four objects would be CLAGN.

\section{Discussion}
In this section, we examine the statistics from all sources we observed with WiFeS. In both the P2S and our 6dFGS cataloguing effort, sources that have both SMSS $u$ and $v$ photometry include 72 WiFeS type-1-1.5 AGN and 24 WiFeS type-2-like AGN (we denote this as the OBS sample for the purpose of this discussion). Among these, 22 are Turn-On CLAGN, including 21 from table \ref{MS-TAB} and a known CLAGN NGC\,2617 \citep{ShappeeCL}. Four are Turn-Off CLAGN, with three from table \ref{MS-TAB} and the known CLAGN MRK\,1018 \citep{mcelroy, husemann}. We do not consider J0014181 as a Turn-Off CLAGN in this analysis because it was not selected with a type-1-like spectrum in 6dFGS. We consider only the earliest type-1.5 spectrum for J0917272 as it is the closest to the SR epoch. Similarly we only consider the type-1.9 spectrum for J1340153.

\subsection{Purity of type-1-like Seyfert galaxy Selection}\label{disc:purity}
\begin{figure}
    \centering
    \includegraphics[width=0.47\textwidth]{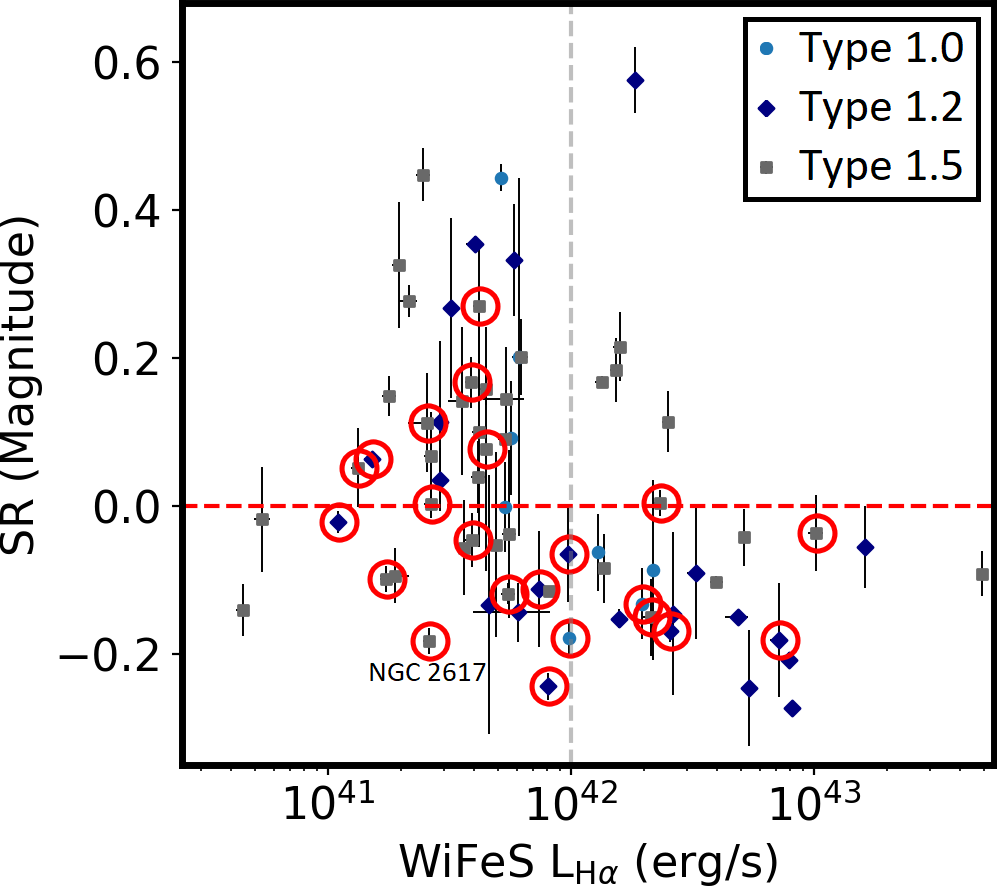}
    \caption{SR versus total broad H$\alpha$ line luminosity from the WiFeS spectrum. Points circled in red are Turn-On CLAGN. Uncertainties for SR error bars are derived from half ranges of $u$ and $v$ photometry measurement. Uncertainties for line ratio error bars are derived from bootstrapping the line fitting measurements. The vertical grey dashed line indicates the separation of high and low luminosity AGN at 10$^{42}$ erg/s as discussed in the section \ref{disc:on-success}. Horizontal red dashed line indicates SR value of 0.}
    \label{disc:fig:1sel}
\end{figure}

\begin{figure*}
    \centering
    \includegraphics[width=0.75\textwidth]{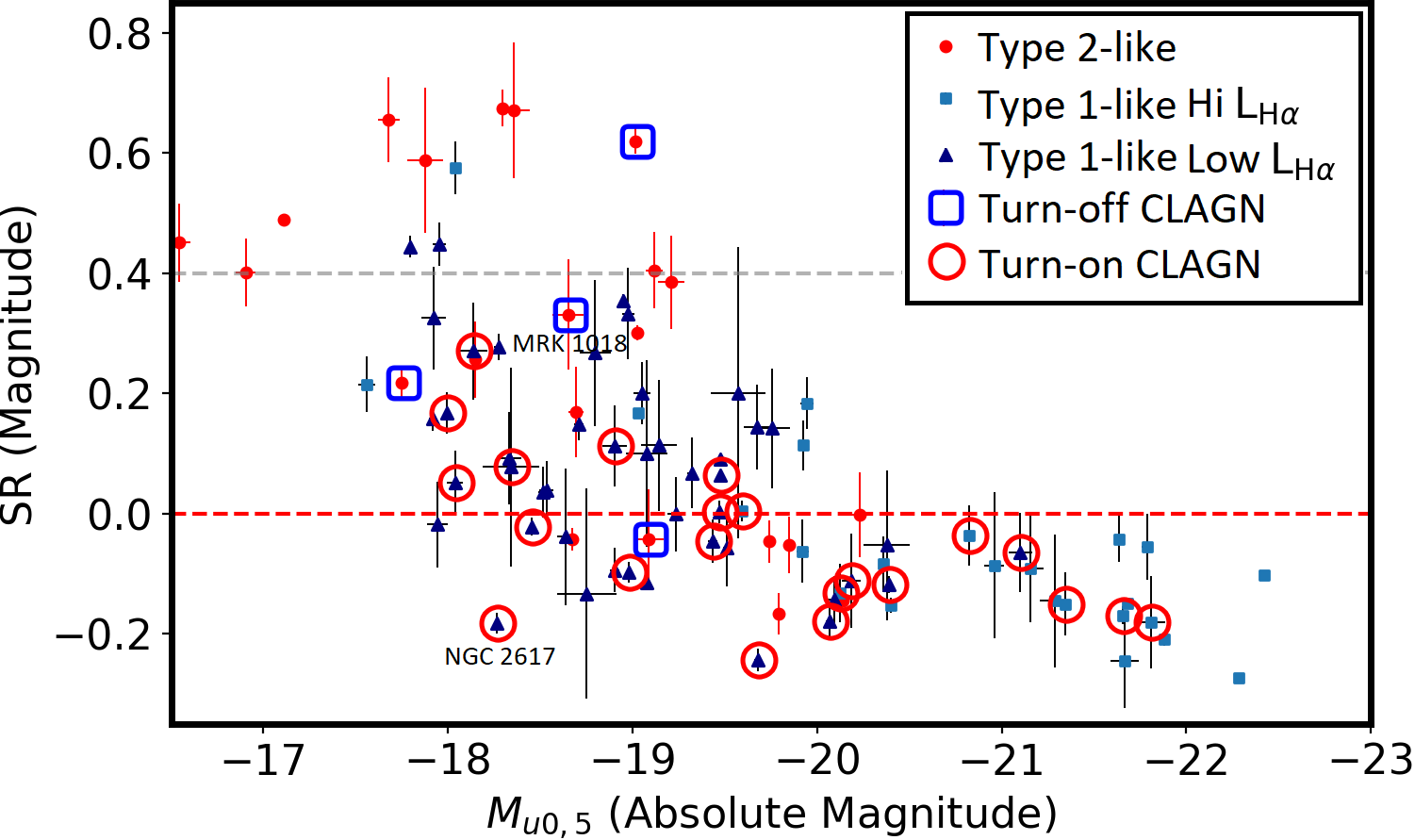}
    \caption{SR magnitude versus $u$-band Absolute Magnitude for the AGN in the OBS sample, including 25 of the CLAGN. Uncertainties for all error bars are derived from the half ranges of $u$ and $v$ photometry measurement. The grey dashed line at SR$=0.4$ indicates the potential for a new selection threshold for Turn-Off CLAGN candidate.}
    \label{disc:fig:12dist}
\end{figure*}

In Paper~I, the SR$=0$ cut was stated to select $\sim75\%$ of type-1 to type-1.5 AGN\footnote{The cut also selected 5 of 6 BL~Lac objects in the parent sample.}, $20\%$ of type-1.8 to type-2 AGN and $\sim7\%$ of star-forming sources. These purity statistics were based on sources within 6dFGS cross-matched with the Hamburg-ESO survey \citep[HESQSO;][]{Wisotzki00} and the Siding Spring Southern Seyfert Spectroscopic Snapshot Survey \citep[S7;][]{Dopita15, Thomas17}. The estimated purity of the SR$=0$ cut explicitly excluded Turn-Off CLAGN that have transitioned in the time since the spectra were obtained (as much as 2 decades earlier), and so could potentially be even higher as an instantaneous measure.

We compare the purity of our selection rule from Paper~I to that obtain from the OBS sample, since the latter now has updated spectral typing acquired from WiFeS. The cut selected 38 out of 72 ($53\%$) from the type-1 to 1.5 AGN sample, and 7 out of 24 ($29\%$) of type-1.8 to 2 AGN. These values are significantly different from that of Paper~I, especially for type-1 to 1.5 AGN. HESQSO is an unbiased catalogue of type-1 AGN, while the S7 survey is close to an accurate representation of overall AGN population. However, the 6dFGS survey was designed to target extended sources by selecting galaxies with high surface brightness or by visual inspection. This leads to a bias towards Seyfert galaxies with mixing ratios that are host galaxy-favoured. Since mixing ratio is affected by AGN luminosity, we account for this bias by considering the distribution of AGN luminosity among the OBS sample. 

In figure \ref{disc:fig:1sel} we plot the SR magnitude as a function of WiFeS H$\alpha$ luminosity. Based on the distribution we observe on the figure, we can separate the type-1-like sample at $L_{H\alpha} = 10^{42} \mathrm{erg/s}$. We then have 19 out of 25 ($76\%$) AGN selected at high luminosity, which is consistent with the statistics presented in Paper~I, while 19 out of 47 ($40\%$) AGN are selected at low luminosity. 

The statistics of the type-2-like AGN are harder to compare since the sample size of our OBS sample and that used in Paper~I are rather small. However, we note that our OBS sample includes the serendipitous CLAGN J1839358 and another non-CLAGN. We suspect that the WiFeS observations have missed the On-state of both of these AGN, while the SMSS photometry did not (see discussion in section \ref{res:midir} and appendix). If we remove these two from the OBS sample, then we only have 5 out of 22 ($23\%$) AGN selected, which is much closer to the value presented in Paper~I.

\subsection{Turn-On CLAGN success rate and selection}\label{disc:on-success}
We plot the SR values, $L_{H\alpha}$, and absolute $u_{0,5''}$ magnitude of the 22 CLAGN within the OBS sample figure \ref{disc:fig:1sel} and \ref{disc:fig:12dist}. J0248345 is excluded as we only observed changes between type-2 and type-1.9 and so it is not comparable to the other Turn-On CLAGN.

The success rate for the SR cut method in selecting Turn-On CLAGN is 13 out of 18 P2S Turn-On candidates ($72\%$). This is a promising rate as it indicates that our search method for CLAGN is the most successful to date. In addition, 16 Turn-On CLAGN ($72\%$) have $L_{H\alpha} < 10^{42} \mathrm{erg/s}$. We discovered these CLAGN by selecting candidates (P2S sample) or observing AGN (serendipitous CLAGN) with 6dFGS spectra of type-1.8, 1.9 or 2. We are therefore blind to the On-state H$\alpha$ luminosity and mixing ratio. Our results show that the CLAGN properties settles into a distribution with more occupying the low luminosity region, and more low luminosity CLAGN having a redder continuum. This supports the finding from \cite{macleodb} that CLAGN are predominantly low luminosity AGN, or AGN with smaller black hole masses.

The majority of the serendipitous CLAGN that our search method missed are located in the low luminosity section. These are J0010100, J0458403, J0519358, J0612386, J1538448, J1721342 and J1958565. We mentioned that these either have mixing ratios that favour the host galaxy, or are potentially CLAGN with fading BELs. 

As mentioned in section \ref{disc:purity}, the low luminosity regime is where our search method is least effective, as the selection rate here is less than $50\%$ (and the selection cut was not design for AGN with these properties). As such, the only way to uncover Turn-On CLAGN in this region would be to monitor all type-2-like AGN frequently, which is currently underway. 

In the high luminosity section, the only serendipitous CLAGN that our search method missed is J1949137 (see section \ref{res:midir}). We believe the SMSS photometry measurement was taken slightly before the Changing-look event occurred, leading to a SR value that is close to 0. 

Overall, our CLAGN search method has a bias towards high luminosity AGN. Only 9 out of 16 ($56\%)$ low luminosity CLAGN are selected, while 5 out of 6 ($83\%$) high luminosity CLAGN are selected. Since this bias is largely attributed to the mixing ratio of the Seyfert galaxy, the only way to reduce this bias is by using smaller aperture photometry, although it will not eliminate the bias completely. That said, since there are more low luminosity CLAGN, being able to select $>50\%$ of them systematically is a promising result in the context of CLAGN searches, as there are no systematic searches among low luminosity AGN.

\subsection{Turn-Off CLAGN selection}\label{disc:off-success}
As mentioned in section \ref{method:srcut}, the purpose of the P2S Turn-Off CLAGN sample is to determine the cut-off required in SR to select Turn-Off CLAGN candidates, and it is therefore expected to have a low probability of recovering any CLAGN. We have discovered 2 Turn-Off CLAGN in P2S among 24 ($8\%$) candidates, while also recovering MRK\,1018. 

The low number of Turn-Off CLAGN found indicates that this method is inefficient at finding these types of CLAGN. That is to be expected, since the SR method was derived only to select AGN from galaxies with high purity. It will not be able to distinguish between type-1 AGN with mixing ratios that are galaxy-favoured and actual type-2 AGN. This is why the region above SR$>0$ is still highly populated by type-1-like AGN. While there are still avenues to explore, such as the region above SR=0.4 with only two type-1-like AGN, a more efficient method for Turn-Off CLAGN will be required in future surveys.

\subsection{Potential TDE or SNe among our CLAGN?}
Tidal Disruption Events (TDE) and Type IIn supernovae (SNe) are two types of events that, near the peak phase in their brightness evolution can look similar spectroscopically to CLAGN \citep[e.g.,][]{tde2, tde3, tde4, sne1}. The main feature that distinguishes between the 3 classes is the duration of the changes. The BELs for both TDE and SNe persist for around a year, whereas CLAGN can have BELs that persist for longer. Therefore we will need to monitor our Turn-On CLAGN for at least a year, acquiring multi-wavelength light curves and spectral data in order to confirm that they are indeed CLAGN.

Another feature of TDEs is the appearance of the He BEL as the Balmer BELs fade. The appearance of the He{\sc ii} BEL is seen in J1824525 and J1949137, and the appearance of He{\sc i} BEL is seen in J2330323. J1949137 is unlikely to be a TDE given that the BEL persisted for over two years. The possibility of the other two being caused by a TDE still remains. 

There is also the case of J1109146. This object may not even be an AGN in 6dFGS, but has BEL that, although fading, persisted for over two years. These properties suggest that it might not be any of the three events we are considering here. 

Our main cause for caution here is because the search method is unable to distinguish between these three events. The method relies on SMSS $u-v$ colour and was designed by considering the clear separation of colours between a luminous QSO and a galaxy. As such, the colour functions more like a switch, being blue for QSO and red for a galaxy. The mixing ratio of a Seyfert galaxy blurs this separation, filling in the gap and resulting in a distribution of colours. This means that the more AGN-dominated the Seyfert galaxy is, the bluer the colour becomes up to a threshold, as the colour saturates to the expected value of a QSO. In the context of Turn-On CLAGN, we are then testing whether the transition to a bluer colour correlates with the appearance of BEL. 

TDE and Type IIn SNe also produce a bluer continuum and the appearance of BEL, and we should expect our search method to select these events as well. It is not trivial to integrate additional constraints to avoid selecting for TDE and SNe \cite[e.g.,][]{tde1, zabludoff21}. Overall, the determination of the causes of the BEL changes have to come after the search, and will require additional information such as high cadence spectroscopy and longer-term light curves.

\subsection{Total CLAGN fraction and population}\label{disc:clagn-frac}
Our results indicate that CLAGN are not a rare phenomena. We have 235 randomly selected AGN in P2S. We also re-observed 90 randomly selected AGN. These samples have large overlaps, leading to a total of 266 randomly selected AGN, of which we have re-observed 132. Among them we have found a total of 26 CLAGN, with 22 Turn-Ons and 4 Turn-Offs. 

18 within the 132 are specifically selected to have a high chance of being a Turn-On CLAGN. Therefore the upper limit to the fraction of Turn-On CLAGN would be $16\%$ ($22/132$). The lower limit is $8\%$ ($22/266$). We take the average and estimate it to be $\sim 12\%$.

The Turn-Off fraction should also be similar due to continuity and symmetry, even though this is not demonstrated by our work. Our selection method is not effective at finding Turn-Off CLAGN. In addition 6dFGS does not have a complete sample of type-1 AGN due to being biased towards extended sources and low surface brightness sources. The equal fraction is demonstrated in \cite{RuncoCL} with 11 Turn-Offs and 9 Turn-Ons within 102 SDSS AGN that have H$\beta$ BEL with estimated black hole masses $>10^7M_{\odot}$. 

Given that the average time between 6dFGS and WiFeS spectra is $\sim$15 years, this implies that $24\%$ of AGN would have changed types during this time period. In other words, $15\%$ of AGN will have changed type if observed again 10 years later, or we can expect $1.5\%$ of AGN to change type annually. Hence, CLAGN are certainly not a rarity, which is also demonstrated by the serendipitous CLAGN described in this paper. 

It is currently unknown, whether all AGN are potentially CLAGN and undergo changes in line with the random rate; instead it is also possible that CLAGN are a subclass of AGN with continuously unstable accretion. It is also not certain if the CLAGN fraction would change if we were observing AGN with a different temporal resolution. These open questions will require the use of the next generation surveys to answer conclusively.

\section{Conclusion and Future Work}
We tested a new method to search for Turn-On CLAGN and show preliminary results from a small sample of 235 sources from 6dFGS. We found 13 Turn-On CLAGN out of 18 CLAGN candidates with the selection process presented in Paper~I. The 6dFGS sample and our selection process demonstrated only a small level of bias towards high luminosity AGN and is promising for CLAGN searches among low luminosity AGN. A future paper will present the results of our CLAGN search with the full 6dFGS sample. 

Our photometric selection rule was tuned to be a tool for selecting type-1 AGN with high purity and would thus be unable to distinguish between galaxies with low-luminosity type-1 AGN and actual type-2 AGN. However, while exploring objects not selected by our rule, we also found two Turn-Off CLAGN among 24 observed sources.

We also found 11 new CLAGN serendipitously while obtaining updated spectra of 90 AGN within 6dFGS as part of our efforts to catalogue all AGN within the survey. Together with our P2S sample, this puts the CLAGN population at $\sim24\%$ in $\sim$15 years, or $1.5\%$ annually. Given that CLAGN such as J1340153 exist, with a changing timescale of only 3 months, we believe that the annual CLAGN rate could be higher if we search with finer temporal resolution.

In addition, the variability in WISE mid-infrared magnitudes suggests that the time gap between the SMSS photometry and our WiFeS observations could be too long. CLAGN could have evolved during this period, resulting in different findings and holes in our search method, as is evident by the discovery of serendipitous CLAGN. Narrowing this time gap is certainly beneficial for future studies.

\section{Acknowledgements}
We appreciate the hard work from our WiFeS observing team for their efforts in acquiring the spectra presented in this paper. The team includes the authors of this paper, as well as Patrick Tisserand from the Institut d'Astrophysique de Paris, David Raithel from the Australian National University, and Noura Alonzi, Thomas Behrendt and Aman Chokshi from the School of Physics in the University of Melbourne. 

Parts of this research were supported by the Australian Research Council Centre of Excellence for All Sky Astrophysics in 3 Dimensions (ASTRO 3D), through project number CE170100013. CAO was supported by the Australian Research Council (ARC) through Discovery Project DP190100252.

The national facility capability for SkyMapper has been funded through ARC LIEF grant LE130100104 from the Australian Research Council, awarded to the University of Sydney, the Australian National University, Swinburne University of Technology, the University of Queensland, the University of Western Australia, the University of Melbourne, Curtin University of Technology, Monash University and the Australian Astronomical Observatory. SkyMapper is owned and operated by The Australian National University's Research School of Astronomy and Astrophysics. The survey data were processed and provided by the SkyMapper Team at ANU. The SkyMapper node of the All-Sky Virtual Observatory (ASVO) is hosted at the National Computational Infrastructure (NCI). Development and support of the SkyMapper node of the ASVO has been funded in part by Astronomy Australia Limited (AAL) and the Australian Government through the Commonwealth's Education Investment Fund (EIF) and National Collaborative Research Infrastructure Strategy (NCRIS), particularly the National eResearch Collaboration Tools and Resources (NeCTAR) and the Australian National Data Service Projects (ANDS).

This publication makes use of data products from the Wide-field Infrared Survey Explorer, which is a joint project of the University of California, Los Angeles, and the Jet Propulsion Laboratory/California Institute of Technology, and NEOWISE, which is a project of the Jet Propulsion Laboratory/California Institute of Technology. WISE and NEOWISE are funded by the National Aeronautics and Space Administration.

\section*{Data Availability}
The SMSS data underlying this article are available at the SkyMapper node of the All-Sky Virtual Observatory (ASVO), hosted at the National Computational Infrastructure (NCI) at \url{https://skymapper.anu.edu.au}. The data from SMSS Data Release 3 are currently accessible only to Australia-based researchers and their collaborators. 

\renewcommand{\thefigure}{A\arabic{figure}}
\setcounter{figure}{0}
\section*{Appendix}
\renewcommand{\thesubsection}{A}
\subsection{J1109146-125554 Atypical CLAGN}
This object was selected as a P2S Turn-On candidate with the BPT analysis classifying it as a type-2 AGN. After a closer inspection, we now realise that the BPT result is affected by the lack of H$\beta$ NEL, inflating value of O{\sc iii}/H$\beta$ ratio that is a key aspect in AGN determination. The weak O{\sc iii}, N{\sc ii} and H$\alpha$ NELs along with the lack of H$\beta$ and S{\sc ii} NELs indicates that a narrow line region does not exist in the core of this galaxy, which also has no star formation. 

As the BELs appeared in the ePESSTO and WiFeS spectra, so did the NELs. The O{\sc iii} line appears weaker and rounded in ePESSTO compared to the stronger and sharper appearance in WiFeS. This could be due to the spectral resolution of ePESSTO being lower than that of WiFeS. Regardless, the emerging S{\sc ii} line in WiFeS definitely shows a developing narrow line region. The appearance of NELs along with the BELs is an oddity. Given that the BELs remained, even though weakened, after 2 years between the ePESSTO and WiFeS spectra, it is unlikely that this event is a TDE or Supernova. 

\begin{figure}
    \centering
    \includegraphics[width=0.47\textwidth]{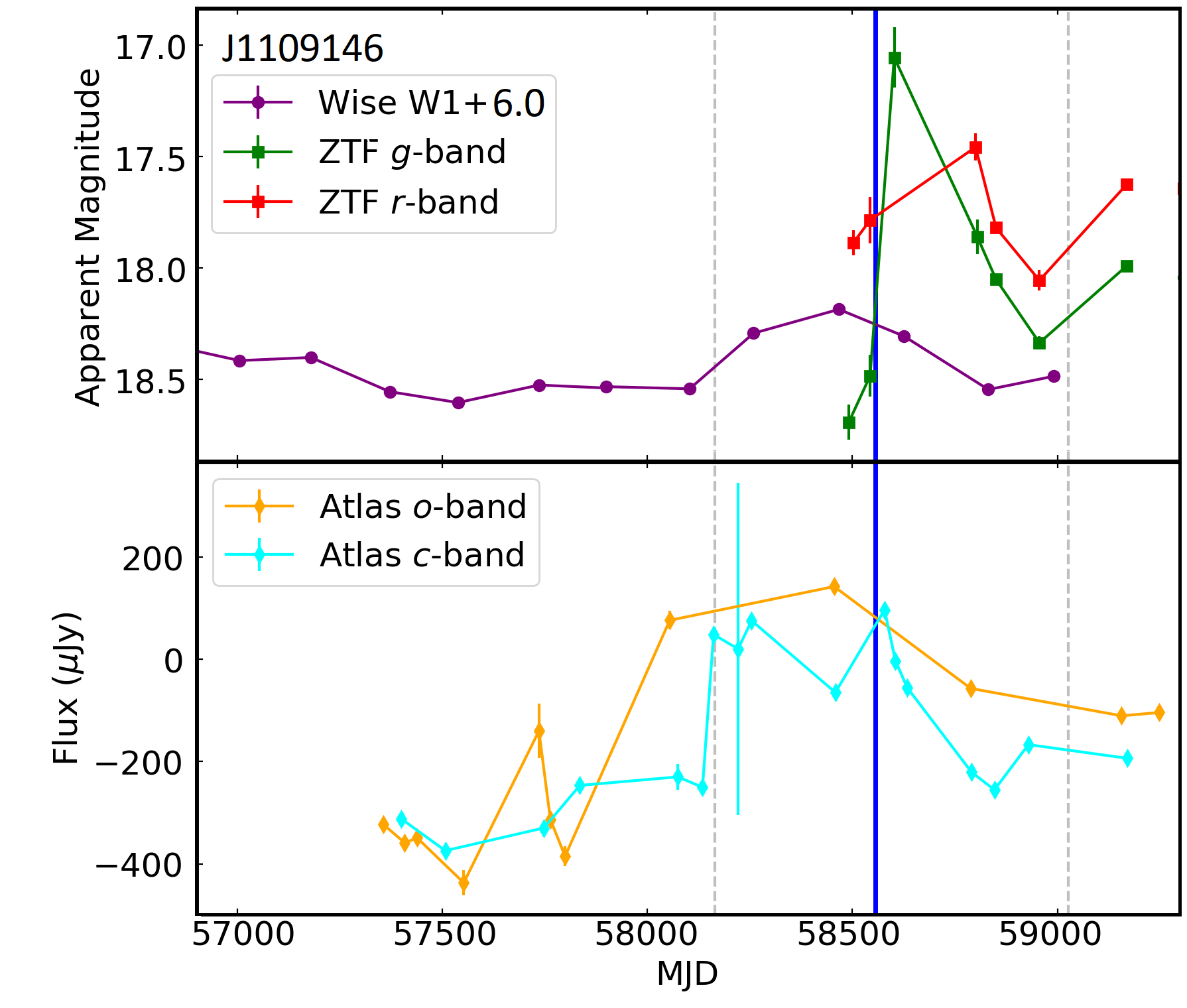}
    \caption{Vertical dotted lines are spectral epoch, with the earlier from ePESSTO and latter from WiFeS. Blue solid vertical line is the epoch of photometry used to calculate the SR magnitude. W1 is offset by 6 magnitude. W1 is binned to 50 days similar to figure \ref{haCL-LC}, but ATLAS and ZTF are binned to 25 days instead.}
    \label{apdx:J1109LC}
\end{figure}

The light curve of this object does not reveal much, other than it being a highly variable source. We obtained photometry data from (see figure \ref{apdx:J1109LC}) ZTF, WISE and ATLAS \citep{tonry18}. There seems to be a flare captured by the ZTF $g$-band and ATLAS $c$-band right after the SMSS photometry epoch. While this flare could fit a TDE description according to the ZTF $g$-band, the BELs have formed before this flare. In addition, ATLAS and W1 light curves all suggests that there had been another flare prior to the ePESSTO spectrum. Therefore a TDE event alone cannot fully explain this object. We suspect that this object could be an AGN turning on for the first time, possibly following the idea presented in \cite{becker00}. If so, then UV spectrum should reveal a Fe Low-ionisation Broad Absorption Line QSO (FeLoBAL) where the Mg{\sc ii} BEL should be highly absorbed by the Fe-complex. 

The WISE W1 and ZTF light curves also suggest another flare after the WiFeS spectrum. We will monitor this object closely, and multi-wavelength observations are encouraged.

\renewcommand{\thesubsection}{B}
\subsection{`Missed' CLAGN as predicted from W1 light curves}
In section \ref{res:midir}, J1839358 and J2007513 demonstrated that the evolution of W1, SMSS SR magnitude and the H$\beta$ BEL all correlate for some CLAGN. J0248345 is a type-2 to 1.9 CLAGN that we believe we have missed an On-state, and the WiFeS observation captures the end of the BELs fading away. According to W1 light curve, there was a peak at around MJD=$58000$ and have dimmed by more than 0.4 magnitude by the time we have observed it on WiFeS. 

\begin{figure}
    \centering
    \includegraphics[width=0.47\textwidth]{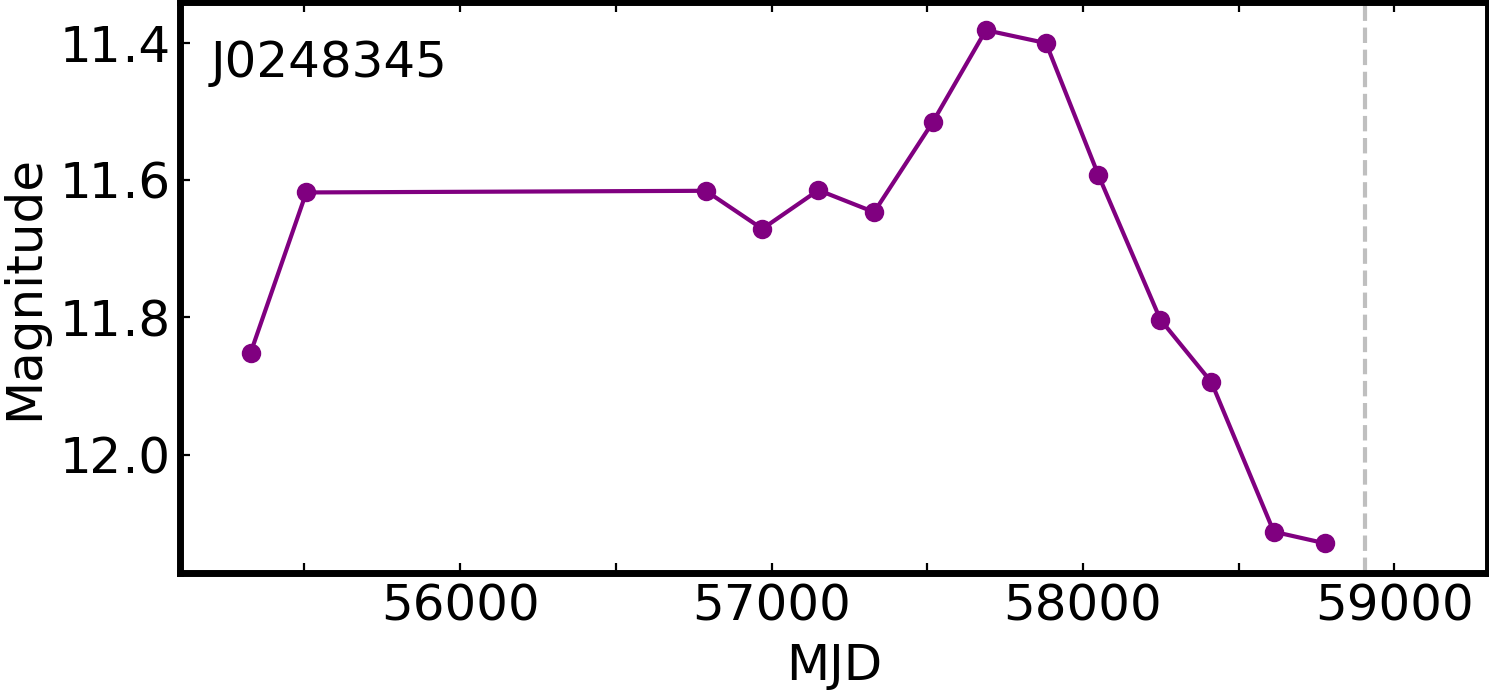}\vspace{-2.5em}
    \includegraphics[width=0.47\textwidth]{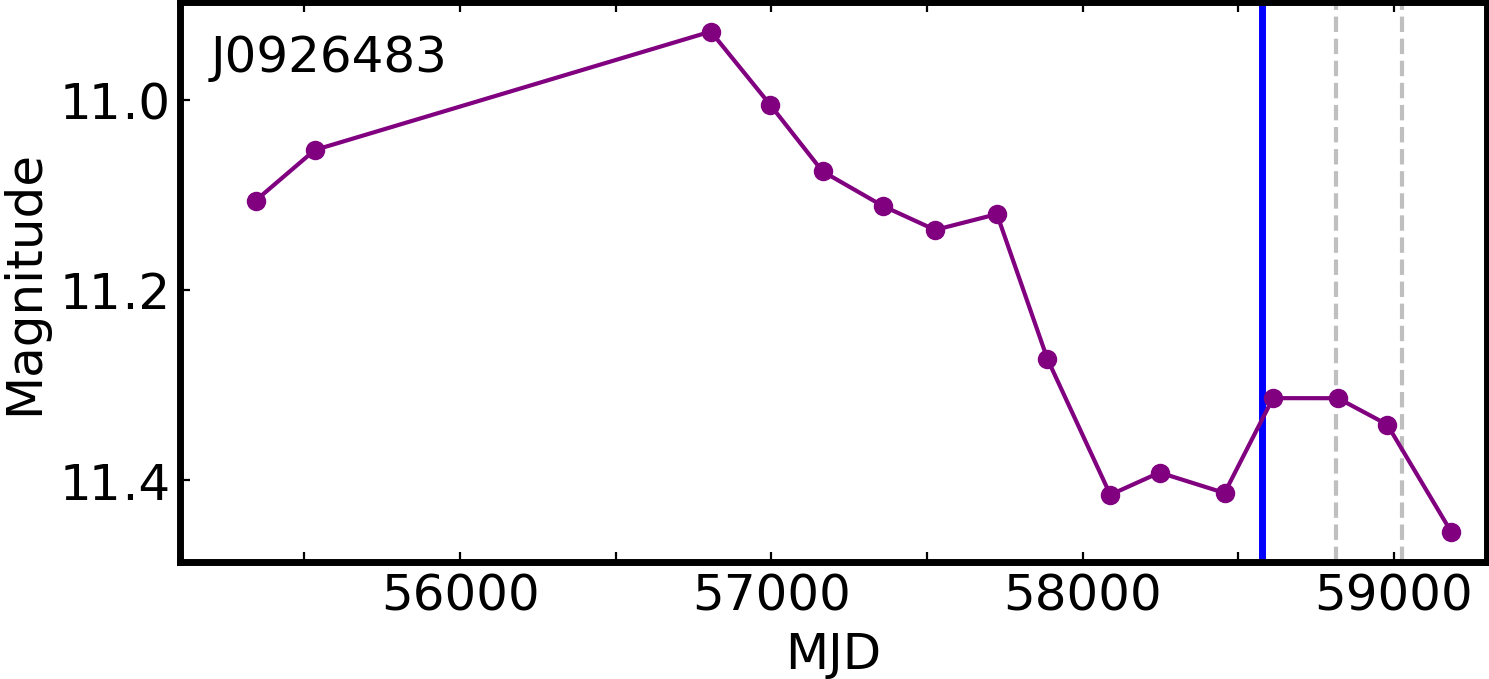}
    \caption{Similar to figure \ref{haCL-LC}}
    \label{adpx:miss}
\end{figure}

J0926483-362608 is not a CLAGN, but has SR=$-0.002$ magnitude. Its spectra is presented at the last panel of figure \ref{BV-APDX2}. Similar to J1839358, the object was much brighter in W1 around MJD=$56000$, and is 0.3 magnitudes or more dimmer during the SMSS and WiFeS observations. There is a slight difference in H$\alpha$ profiles between 6dFGS and WiFeS, suggesting a similar scenario to J0248345 where we have missed the On-state.

\renewcommand{\thesubsection}{C}
\subsection{CLAGN Spectra}
The following figures show all the CLAGN discovered using one or other of the methods describes in the text.
\begin{figure*}
    \centering
    \includegraphics[width=\textwidth]{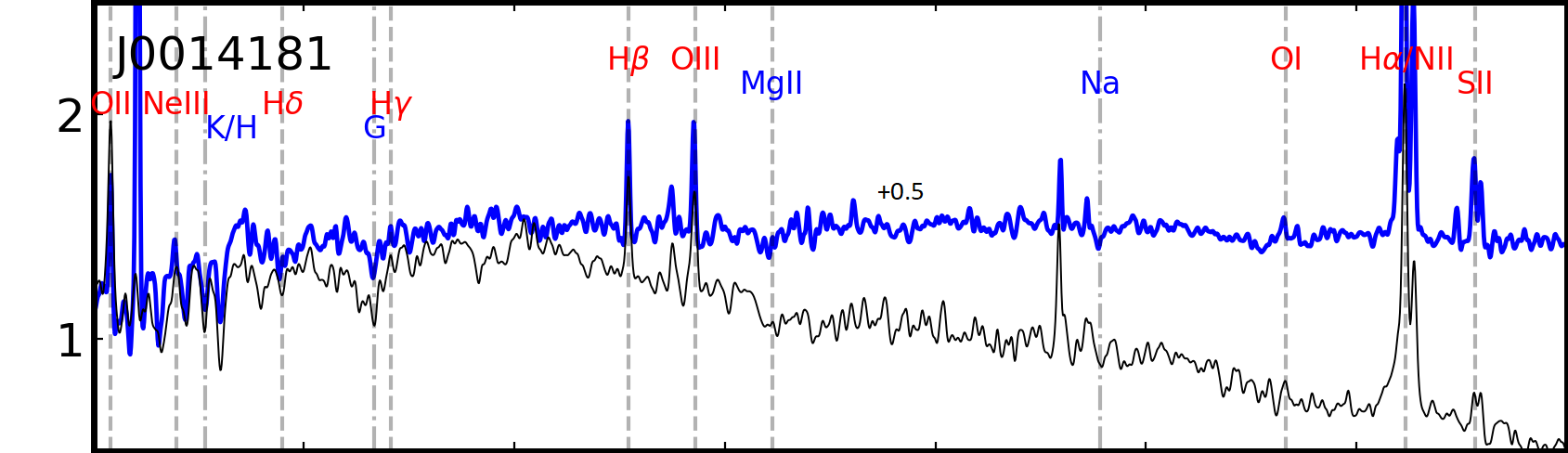}\vspace{-0.2em}
    \includegraphics[width=\textwidth]{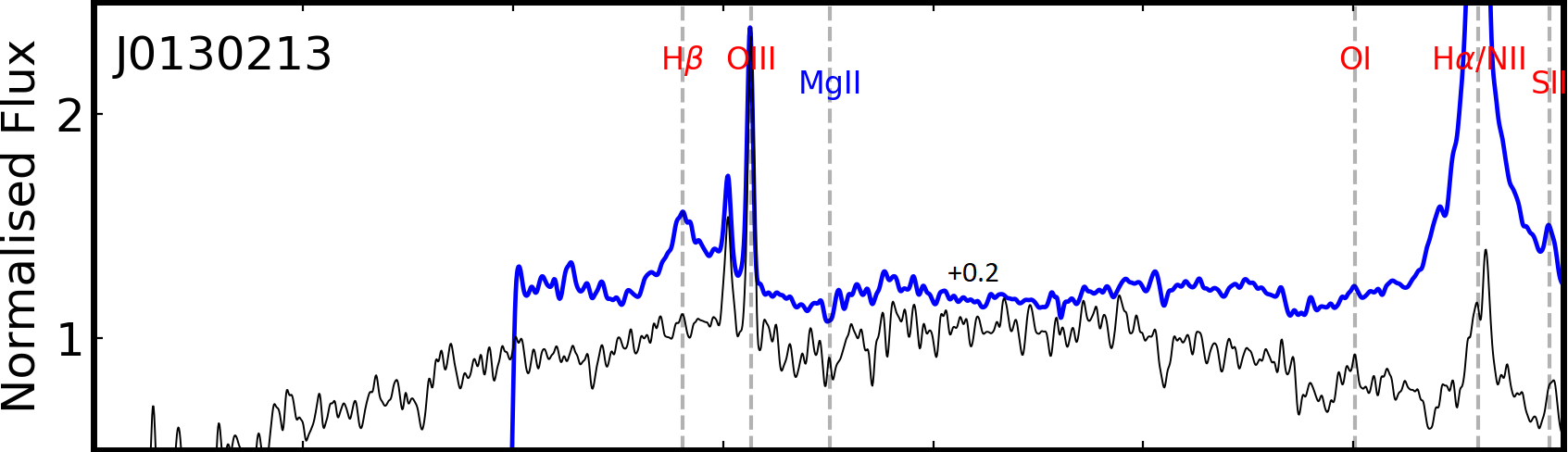}\vspace{-0.2em}
    \includegraphics[width=\textwidth]{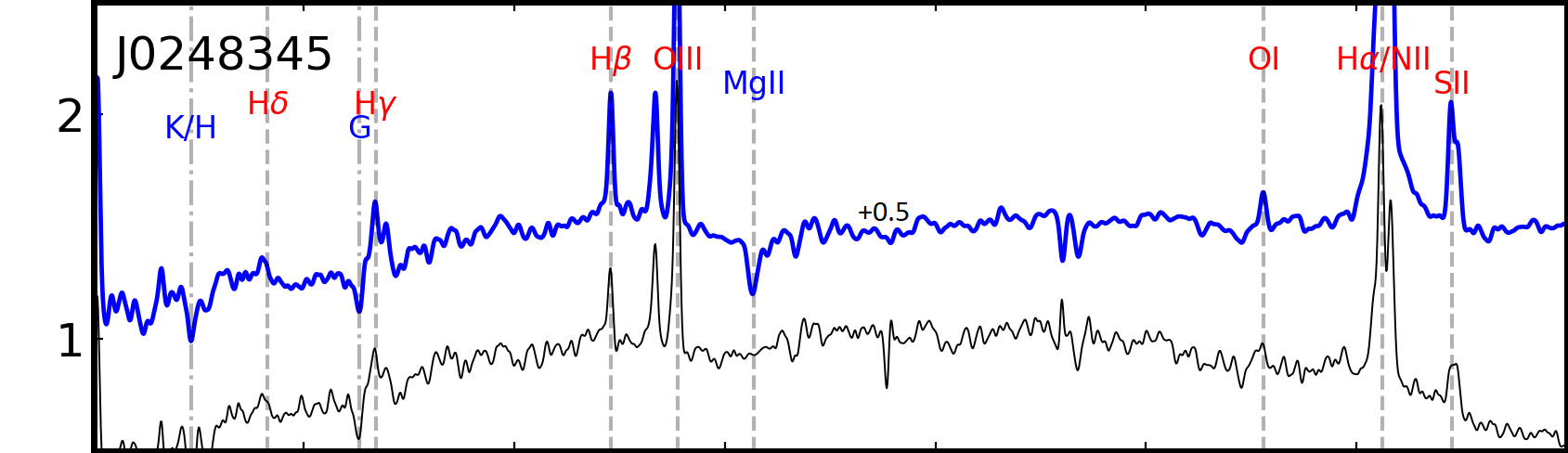}\vspace{-0.2em}
    \includegraphics[width=\textwidth]{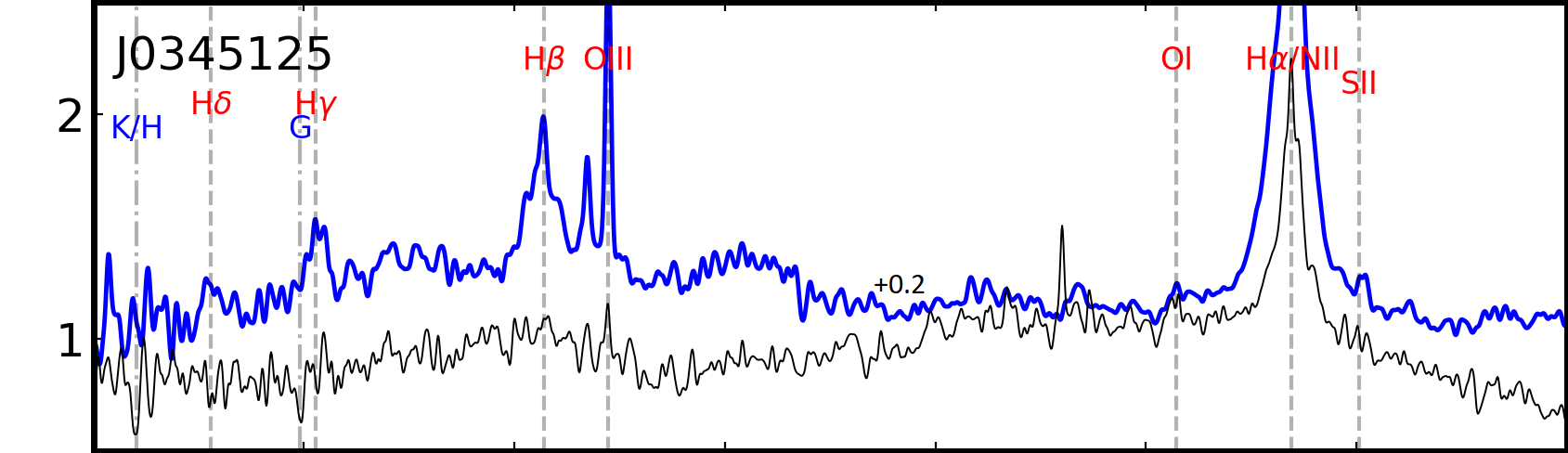}
    \includegraphics[width=\textwidth]{CLAGN-Spec/xaxis-spec.png}
    \caption{Remaining spectra from P2S and Serendipitous sample. Same plot description as Figure \ref{CL-EX1}. J0130213 is taken with RT-480 beam splitter, and the quality of the blue CCD spectrum is too low and does not contain any more information. We chose to not show it here.}
    \label{CL-APDX1}
\end{figure*}
\begin{figure*}
    \centering
    \includegraphics[width=\textwidth]{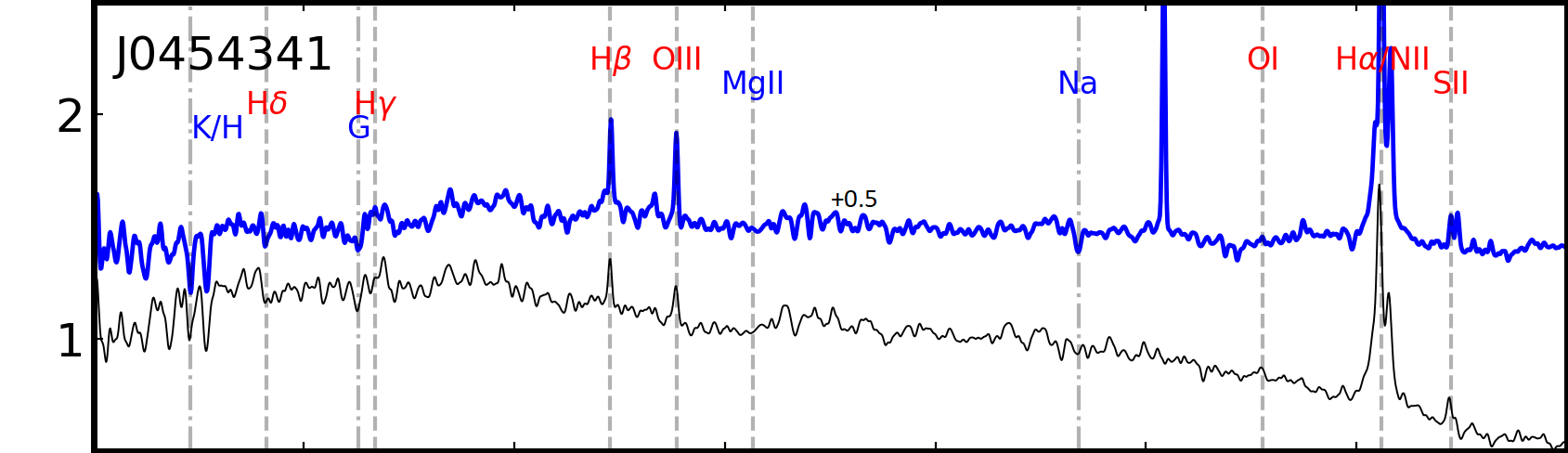}\vspace{-0.2em}
    \includegraphics[width=\textwidth]{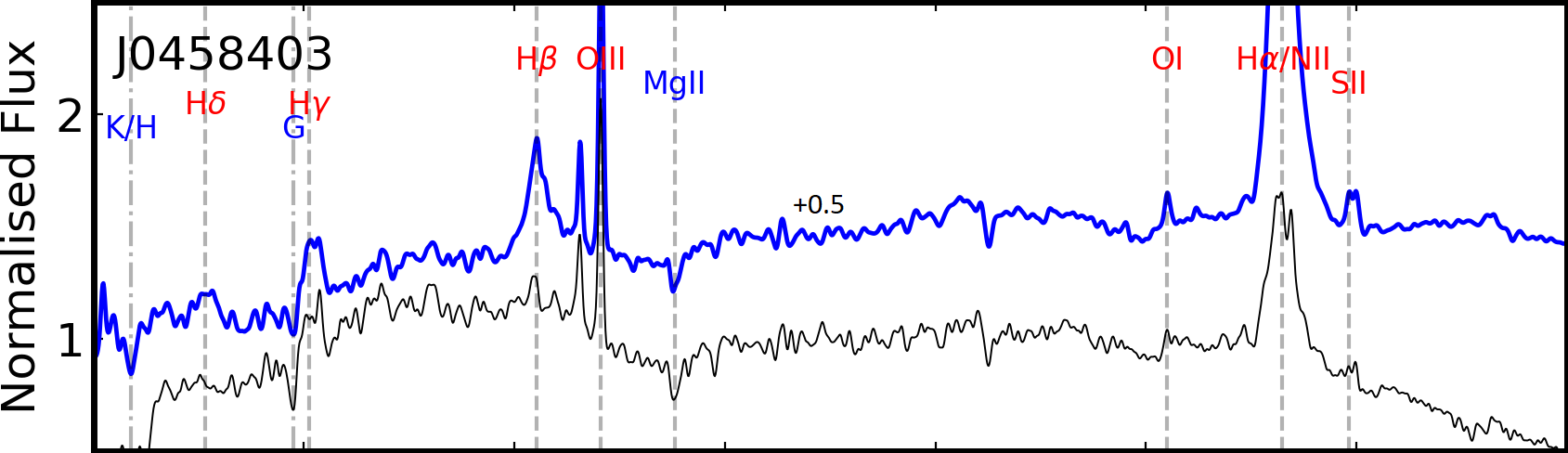}\vspace{-0.2em}
    \includegraphics[width=\textwidth]{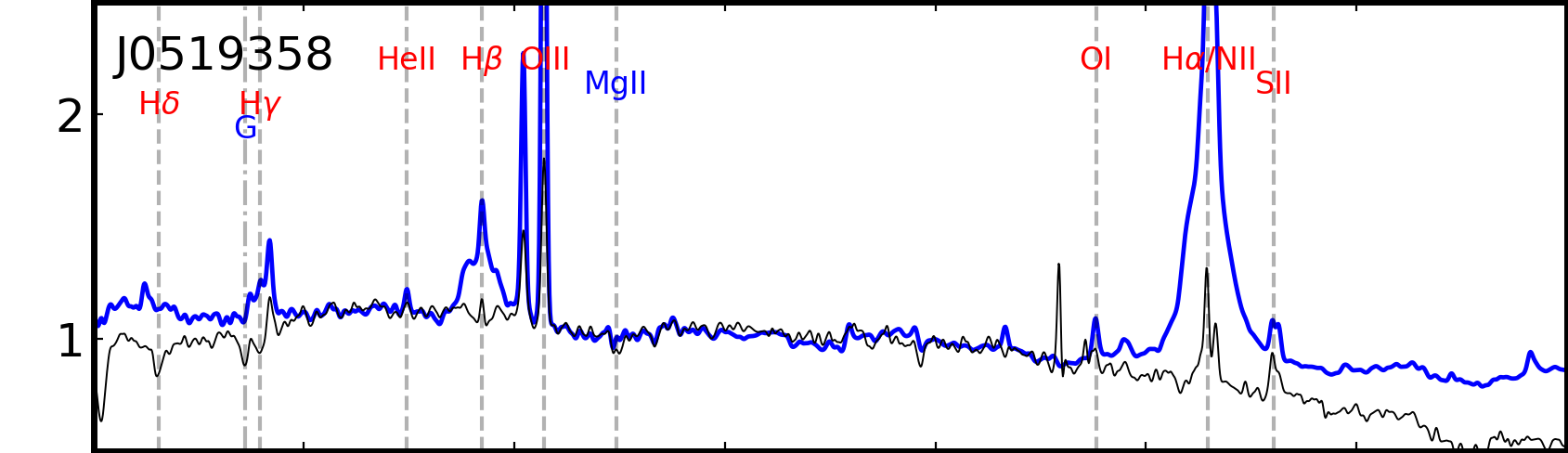}\vspace{-0.2em}
    \includegraphics[width=\textwidth]{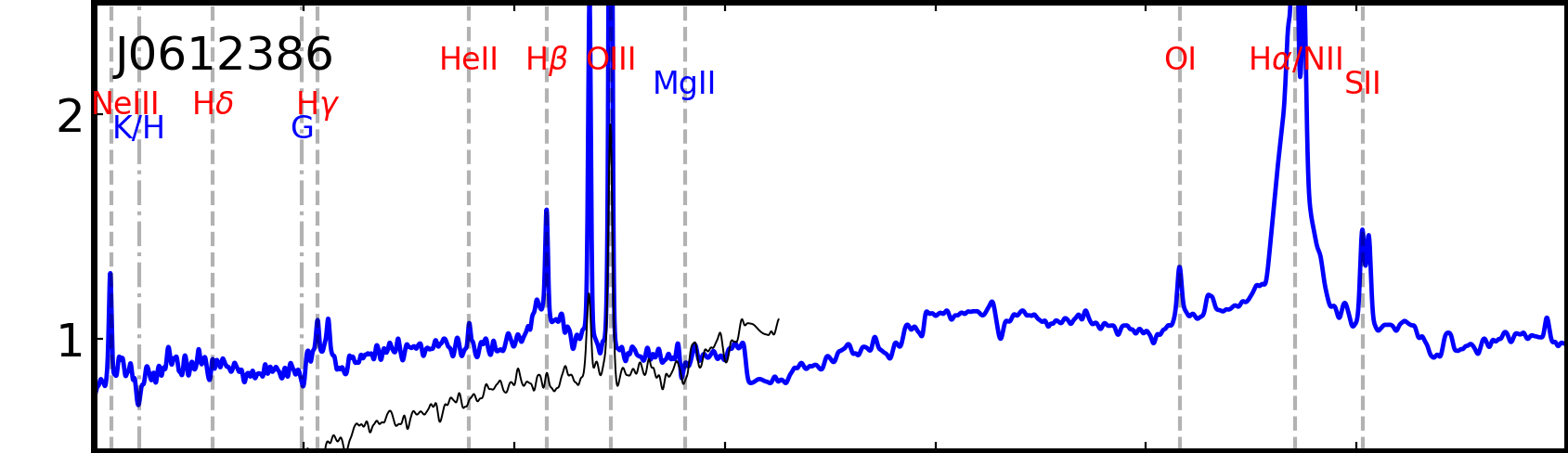}
    \includegraphics[width=\textwidth]{CLAGN-Spec/xaxis-spec.png}
    \caption{Continued from Figure \ref{CL-APDX1}}
    \label{CL-APDX2}
\end{figure*}
\begin{figure*}
    \centering
    \includegraphics[width=\textwidth]{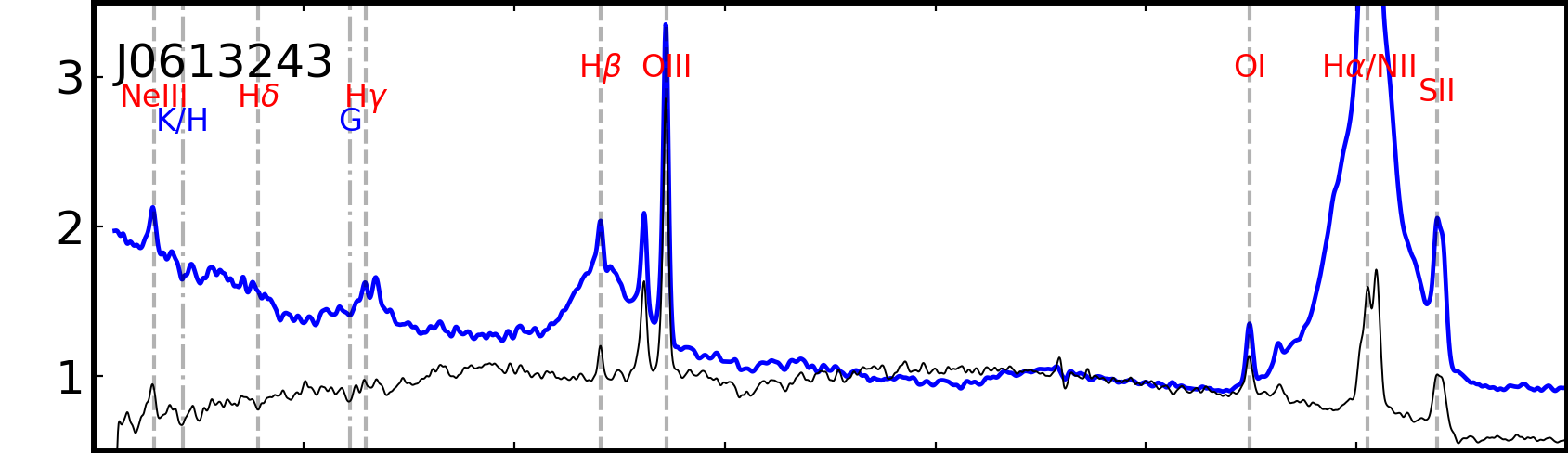}\vspace{-0.2em}
    \includegraphics[width=\textwidth]{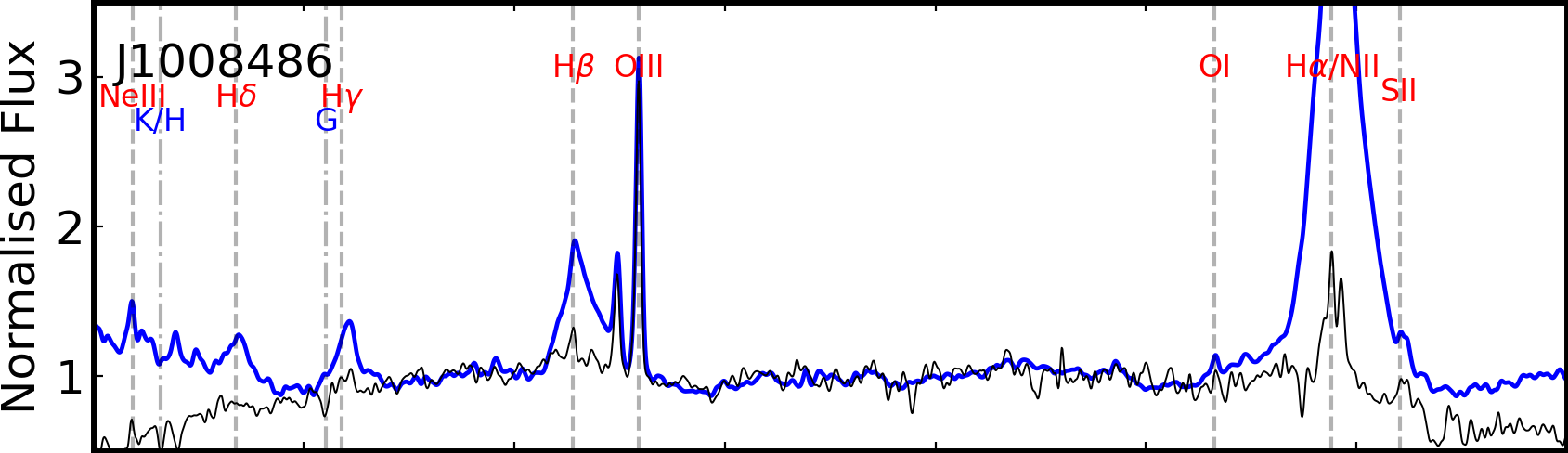}\vspace{-0.2em}
    \includegraphics[width=\textwidth]{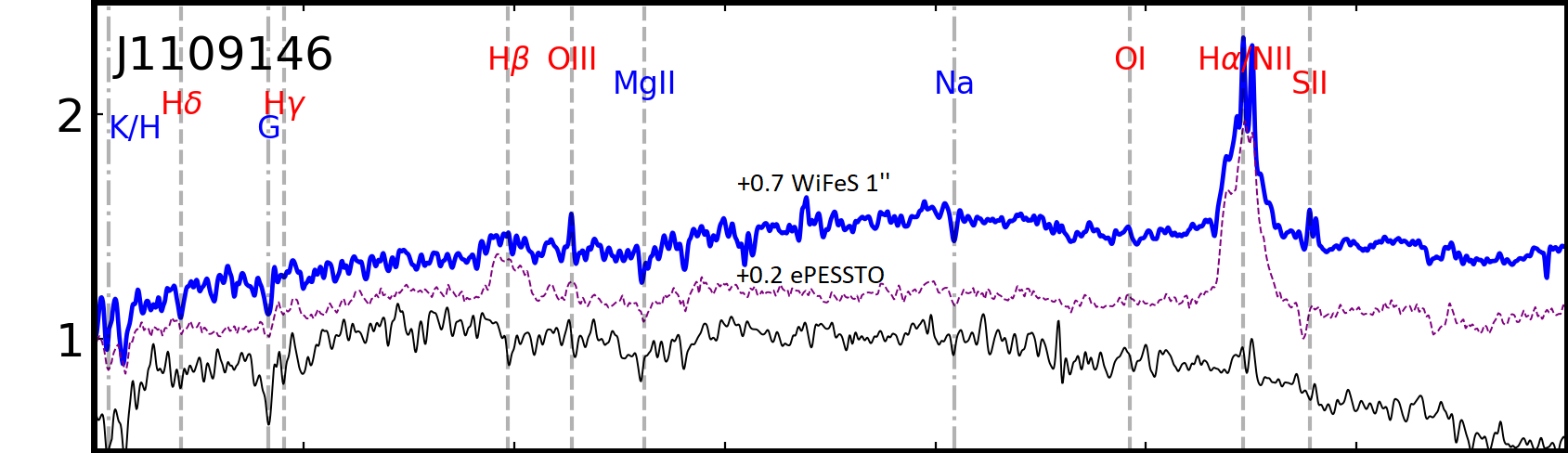}\vspace{-0.2em}
    \includegraphics[width=\textwidth]{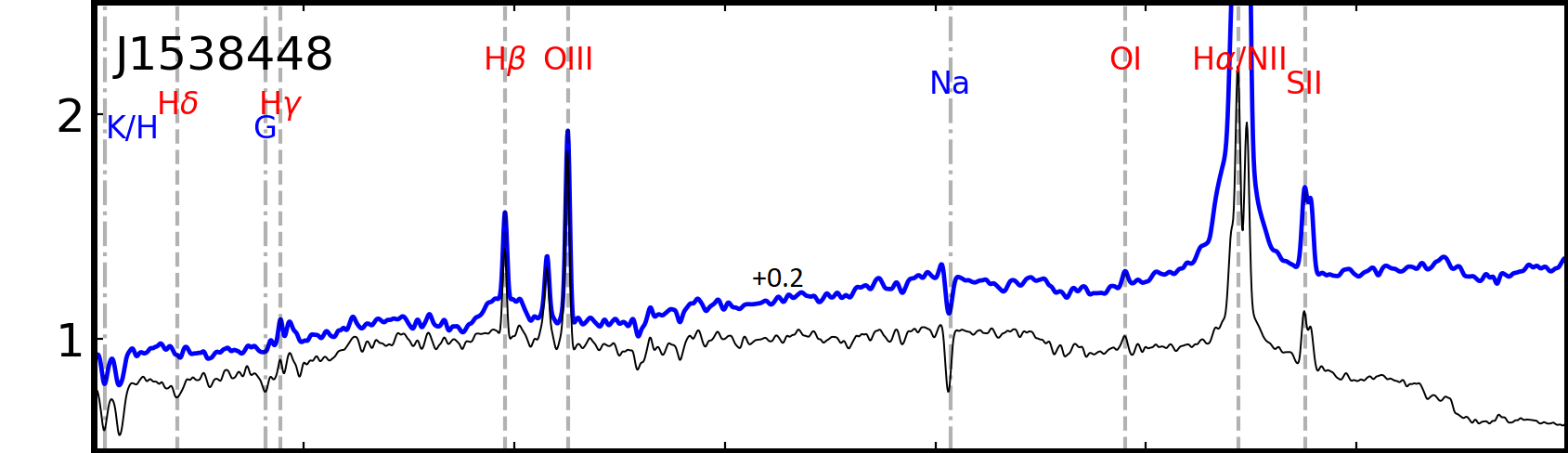}
    \includegraphics[width=\textwidth]{CLAGN-Spec/xaxis-spec.png}
    \caption{Continued from Figure \ref{CL-APDX1}}
    \label{CL-APDX3}
\end{figure*}
\begin{figure*}
    \centering
    \includegraphics[width=\textwidth]{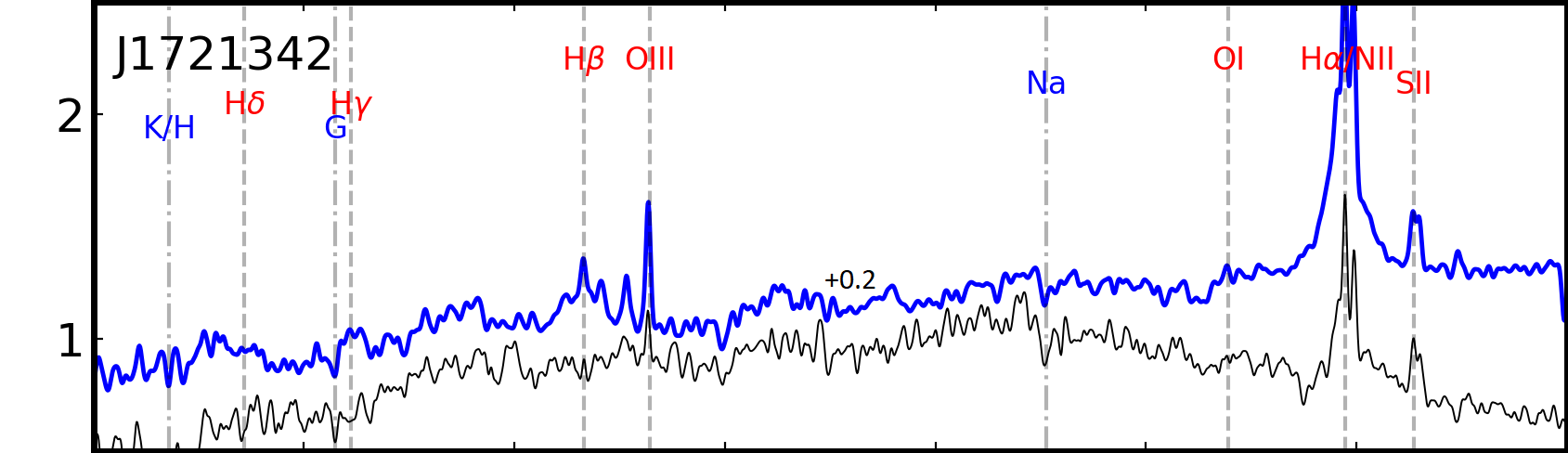}\vspace{-0.2em}
    \includegraphics[width=\textwidth]{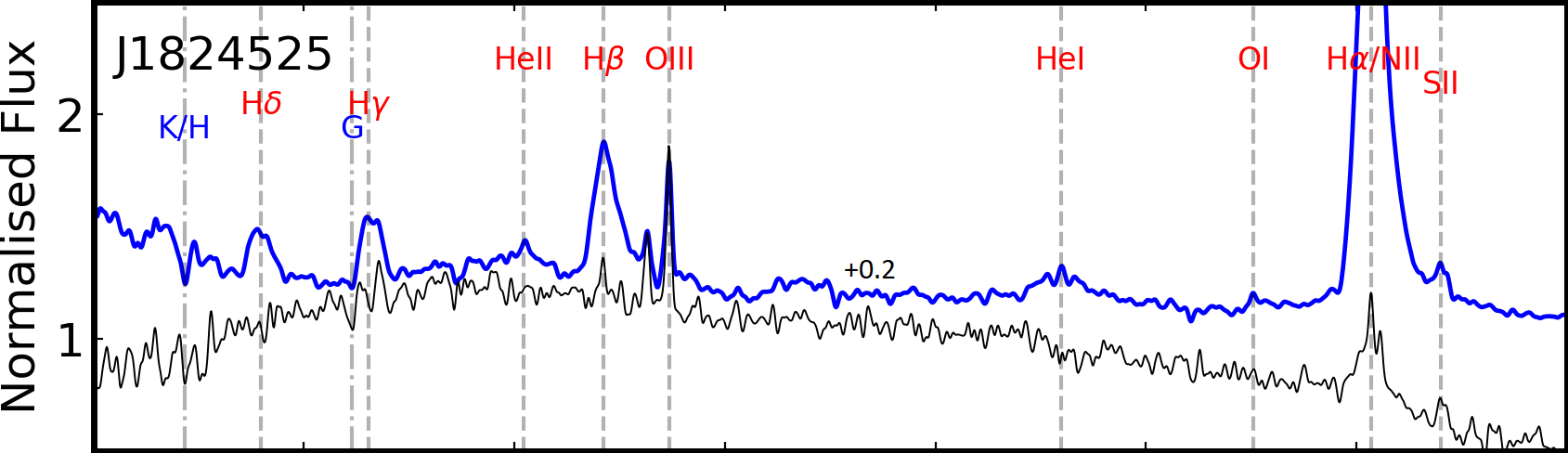}\vspace{-0.2em}
    \includegraphics[width=\textwidth]{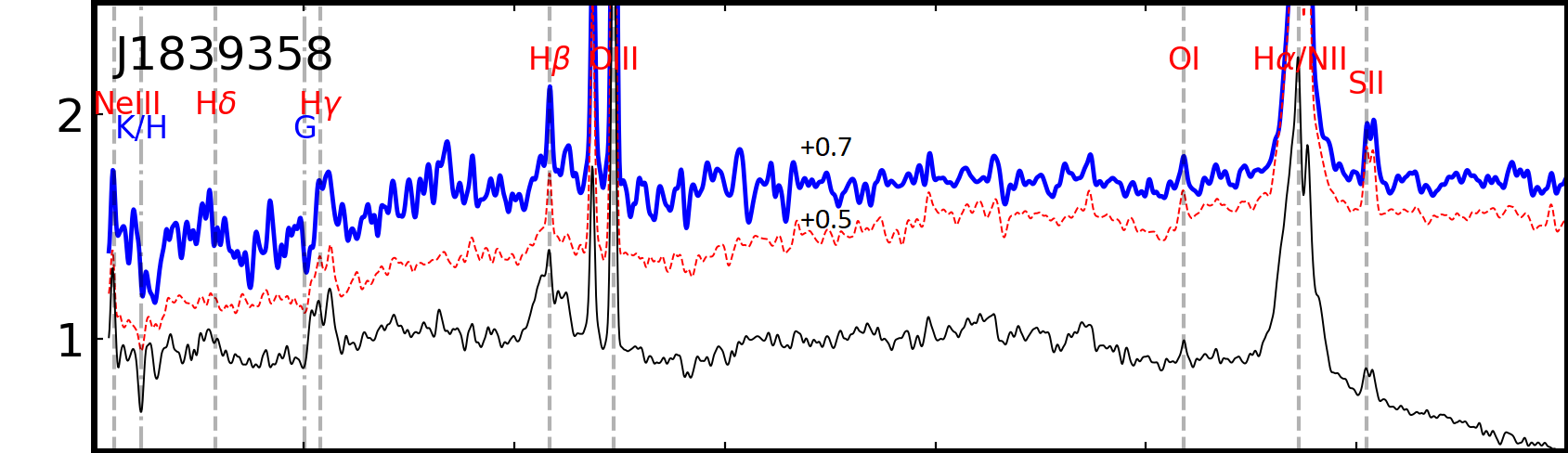}\vspace{-0.2em}
    \includegraphics[width=\textwidth]{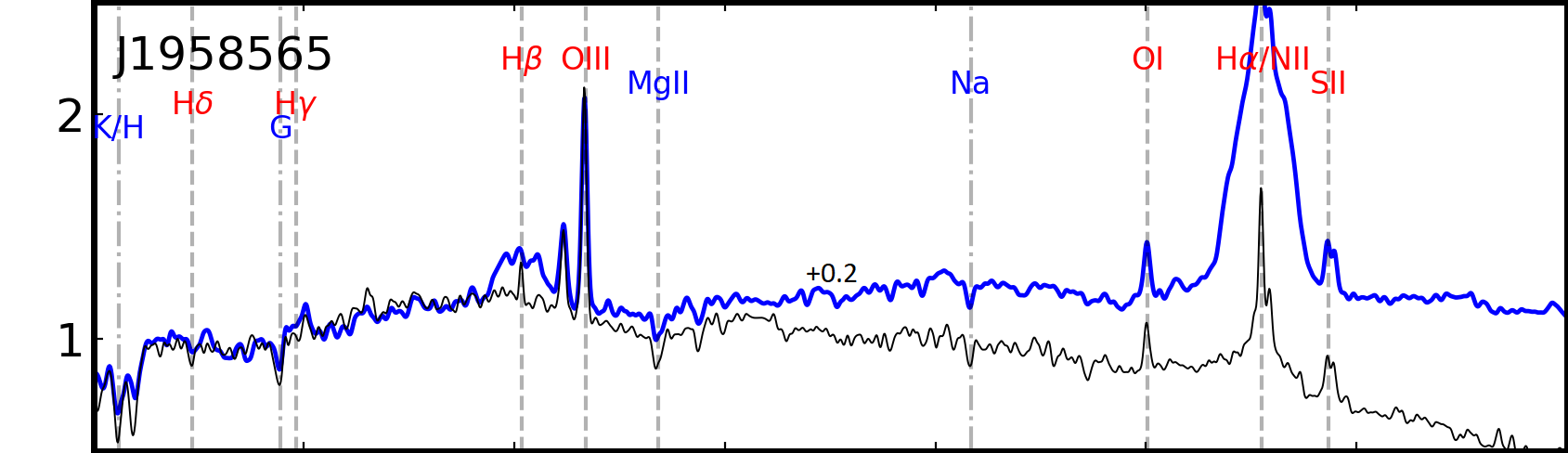}
    \includegraphics[width=\textwidth]{CLAGN-Spec/xaxis-spec.png}
    \caption{Continued from Figure \ref{CL-APDX1}}
    \label{CL-APDX4}
\end{figure*}
\begin{figure*}
    \centering
    \includegraphics[width=\textwidth]{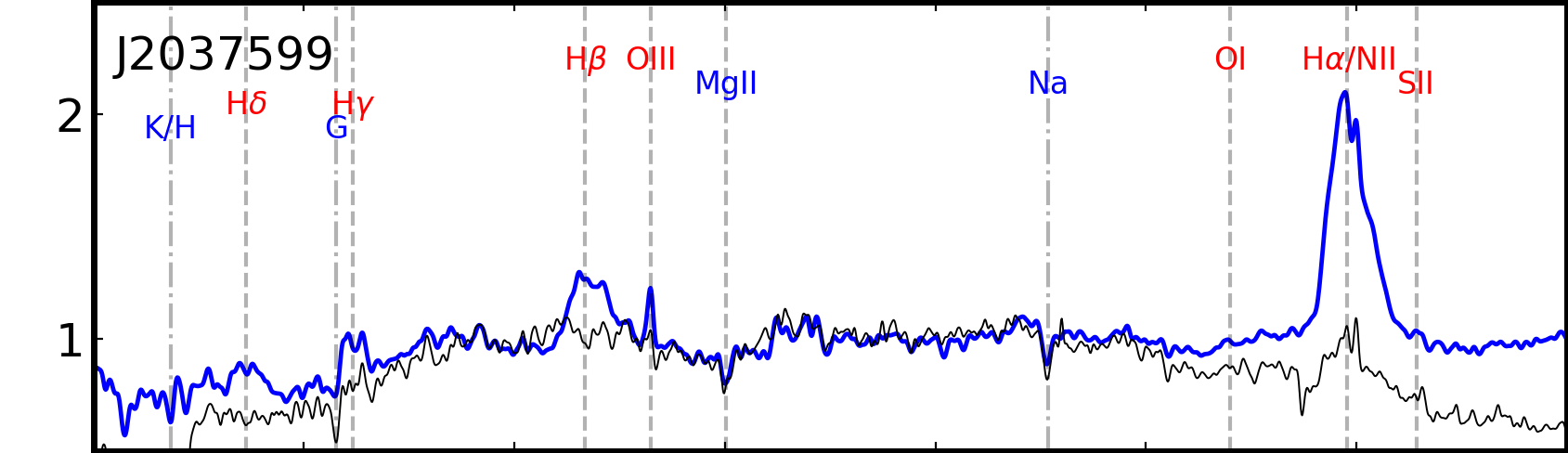}\vspace{-0.2em}
    \includegraphics[width=\textwidth]{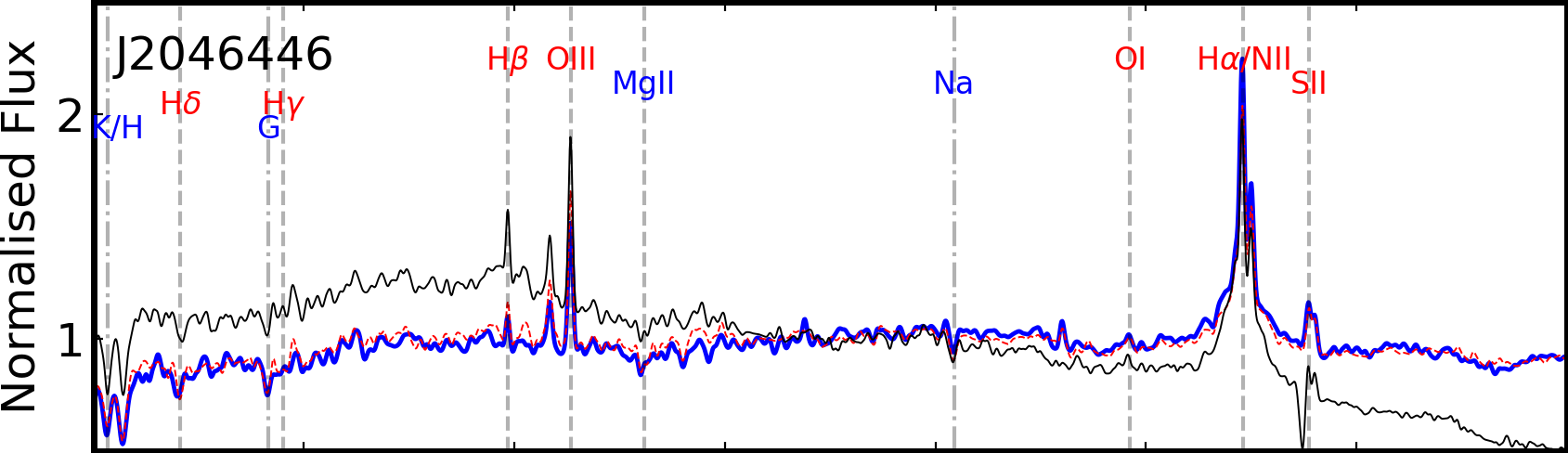}\vspace{-0.2em}
    \includegraphics[width=\textwidth]{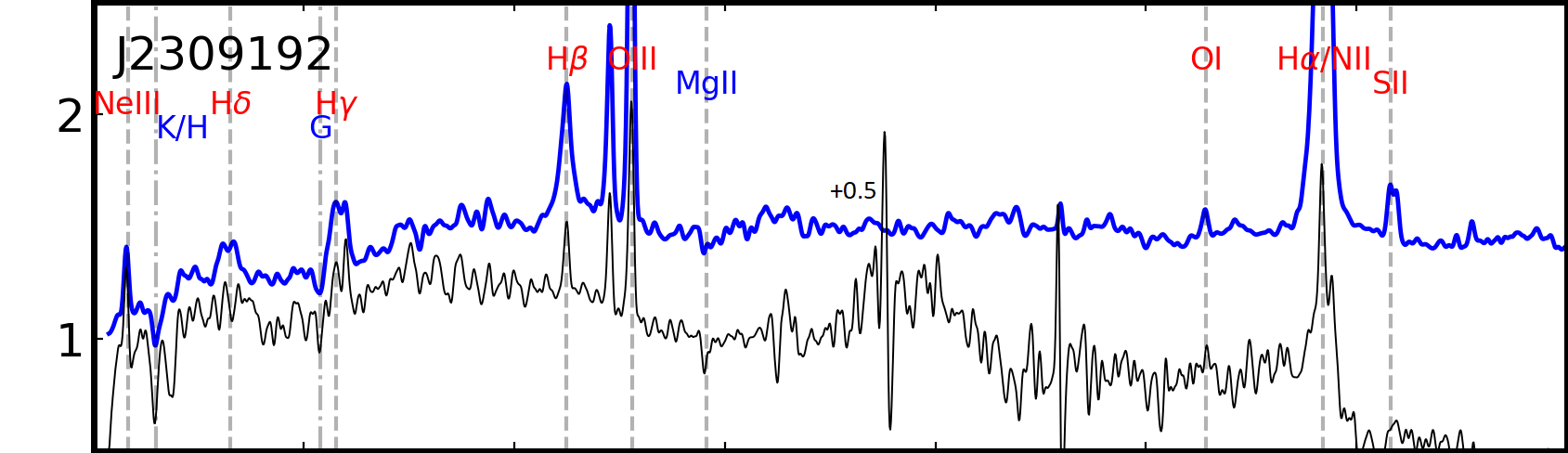}
    \includegraphics[width=\textwidth]{CLAGN-Spec/xaxis-spec.png}
    \caption{Continued from Figure \ref{CL-APDX1}}
    \label{CL-APDX5}
\end{figure*}

\begin{figure*}
    \centering
    \includegraphics[width=\textwidth]{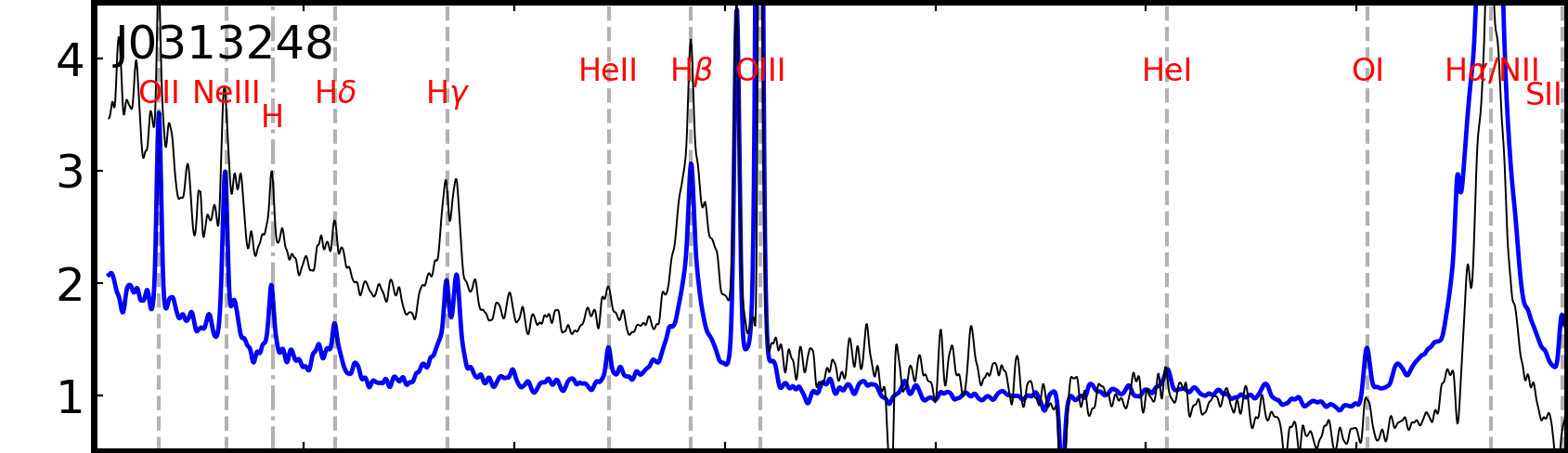}\vspace{-0.2em}
    \includegraphics[width=\textwidth]{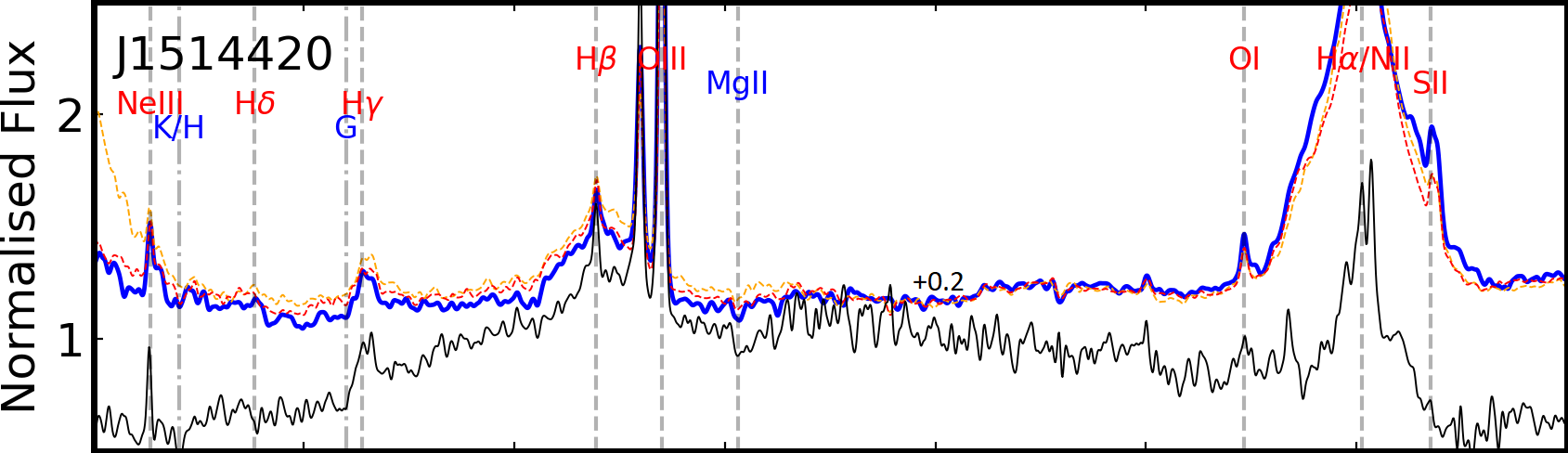}\vspace{-0.2em}
    \includegraphics[width=\textwidth]{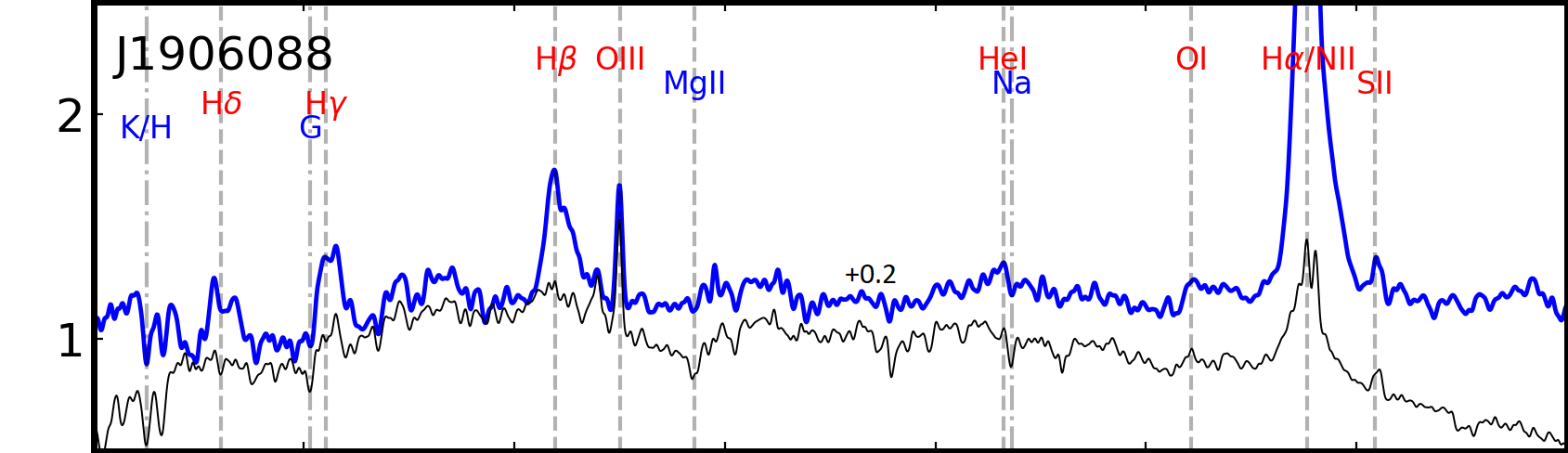}\vspace{-0.2em}
    \includegraphics[width=\textwidth]{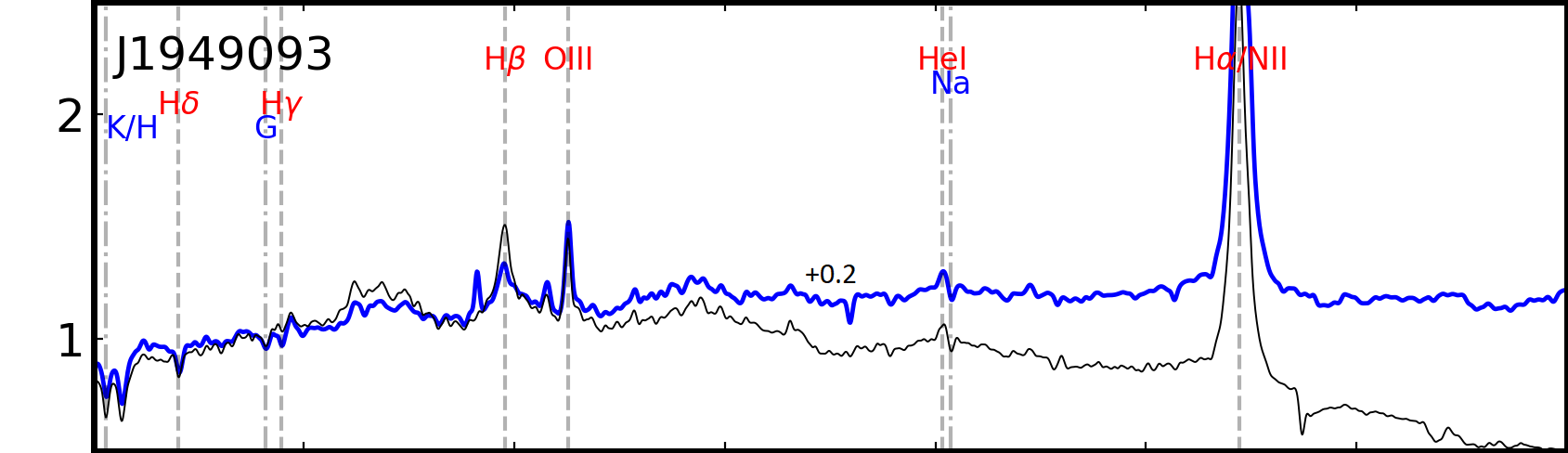}
    \includegraphics[width=\textwidth]{CLAGN-Spec/xaxis-spec.png}
    \caption{Continued from Figure \ref{BV-EX}}
    \label{BV-APDX1}
\end{figure*}
\begin{figure*}
    \centering
    \includegraphics[width=\textwidth]{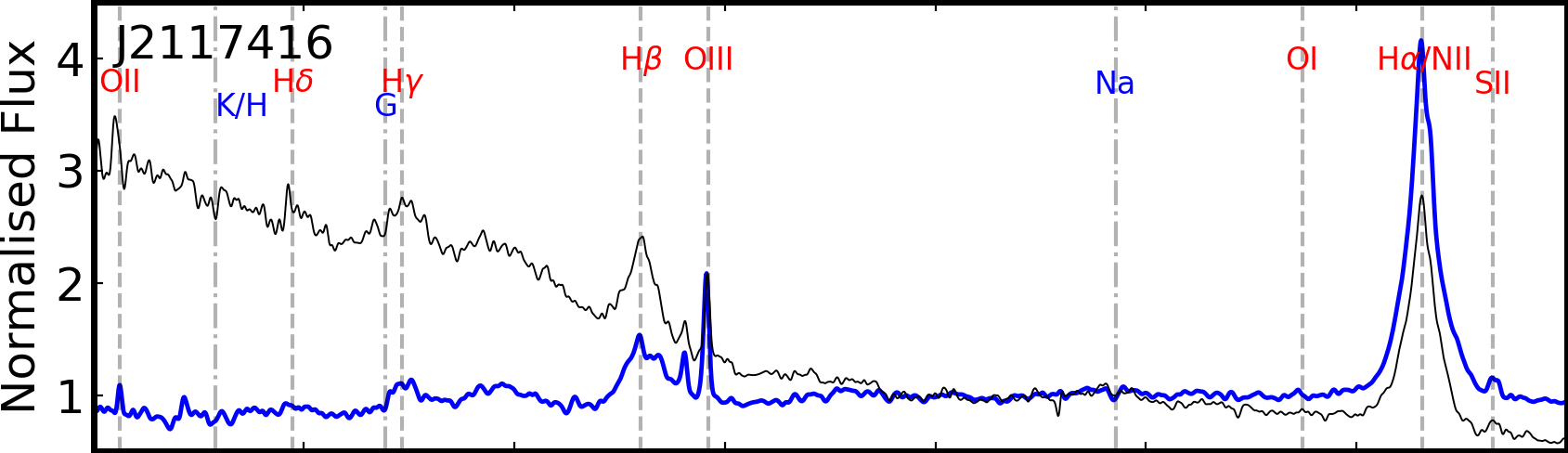}\vspace{-0.2em}
    \includegraphics[width=\textwidth]{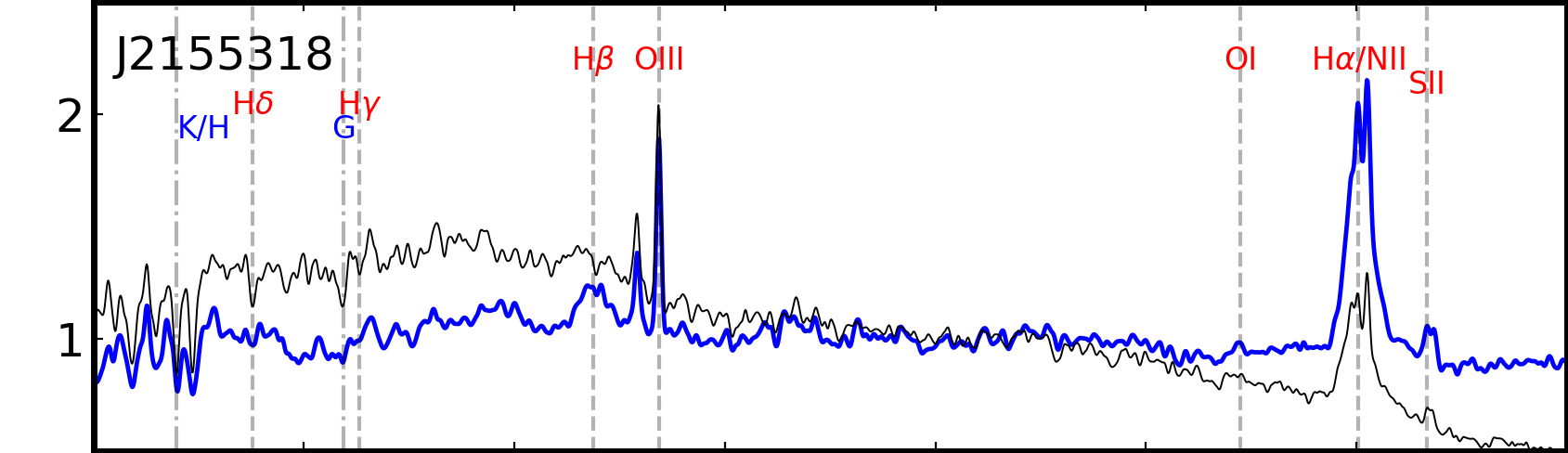}
    \includegraphics[width=\textwidth]{CLAGN-Spec/xaxis-spec.png}\vspace{0.2em}
    \includegraphics[width=\textwidth]{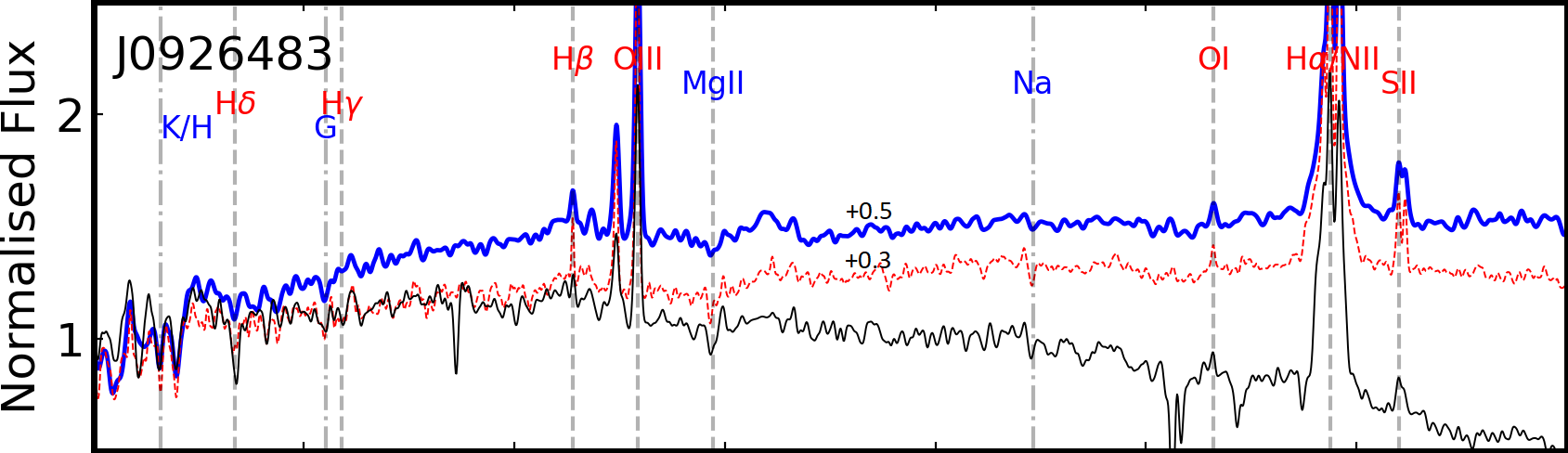}
    \includegraphics[width=\textwidth]{CLAGN-Spec/xaxis-spec.png}
    \caption{Continued from Figure \ref{BV-EX}. The last panel is a `missed' CLAGN from the appendix}
    \label{BV-APDX2}
\end{figure*}

\end{document}